\RequirePackage{lineno}
\documentclass[twocolumn,showpacs,superscriptaddress,amsmath,amssymb,nofootinbib]{revtex4-2}
\usepackage{graphicx}
\usepackage{subfigure}
\usepackage{subcaption}
\usepackage{epsfig}
\usepackage{overpic}
\usepackage{dcolumn}
\usepackage{ulem}
\usepackage{bm}
\usepackage{color}
\usepackage{multirow}
\usepackage{xspace}
\usepackage{epstopdf}
\usepackage{xcolor}
\usepackage{soul}
\usepackage{verbatim}
\usepackage{enumitem}
\usepackage{todonotes}
\usepackage{cancel}
\usepackage[toc,page]{appendix}
\usepackage[colorlinks,allcolors=black]{hyperref}

\usepackage{diagbox}

\newcommand{\ihep}{\affiliation{State Key Laboratory of Nuclear Physics and Technology, Institute of Quantum Matter, South China Normal University, Guangzhou 510006, China
}}

\newcommand{\moe}{\affiliation{Key Laboratory of Atomic and Subatomic Structure and Quantum Control (MOE), Guangdong-Hong Kong Joint Laboratory of Quantum Matter, Guangzhou 510006, China}}

\newcommand{\OU}{\affiliation{Research Center for Nuclear Physics (RCNP), Osaka University, Ibaraki 567-0047, Japan}}

\newcommand{\SUT}{\affiliation{School of Physics and Center of Excellence in High Energy Physics \& Astrophysics, \\ Suranaree University of Technology, Nakhon Ratchasima 30000, Thailand}}

\newcommand{\iqm}{\affiliation{Guangdong Basic Research Center of Excellence for 
Structure and Fundamental Interactions of Matter, Guangdong 
Provincial Key Laboratory of Nuclear Science, Guangzhou 
510006, China}}

\newcommand{\scnt}{\affiliation{Southern Center for Nuclear-Science Theory (SCNT), Institute of Modern Physics, Chinese Academy of Sciences, Huizhou 516000, Guangdong Province, China}}

\newcommand{\gscas}{\affiliation{School of Physics, South China Normal University, Guangzhou 510006, China}}

\begin{document}
\include{def-com}
\title{\boldmath Three-body final state interactions in $B^+\to D\bar{D}K^+$ decays}



\author {Xin-Yue Hu}
\iqm
\moe
\ihep
\author {Jiahao He}
\email{co-first author}
\moe
\gscas
\author {Pengyu Niu}
\email{niupy@m.scnu.edu.cn}
\iqm
\ihep
\author {Qian Wang}
\email{qianwang@m.scnu.edu.cn}
\iqm 
\ihep
\scnt
\OU

\author {Yupeng Yan}
\email{yupeng@sut.ac.th}
\SUT
 
\date{\today}

\begin{abstract}
This paper presents a detailed analysis of the three-body final state  interactions in the $B^+\to D\bar{D}K^+$ process, whose phase space is sufficient small. To precisely extract the resonance parameters, for instance the $\chi_{c0/2}(3930)$ in the $D\bar{D}$ invariant mass distributions, in this process, one has to take into account final state interaction, especially the three-body final state interaction. We employ the dispersive Khuri-Treiman formalism, combined with a parametrization of the $D\bar{D}$ interaction based on heavy quark spin symmetry. By performing a simultaneous fit to the experimental data from LHCb, $BABAR$, and Belle collaborations, the scheme with three-body interaction  successfully describes the invariant mass distributions of the three two-body subsystems.
We precisely extract the pole structures of $\chi_{c0}(3930)$ and $\psi(3770)$ as $3.913-0.016i~\mathrm{GeV}$ and $3.761-0.006i~\mathrm{GeV}$ in $B^+$ decay. By performing the pole trajectory analysis on a uniformized complex plane, we find that both of them stem from the input bare state.
\end{abstract}
\maketitle


\section{Introduction}
\label{sec:Introduction}

In the last two decades, since the discovery of  the $X(3872)$ in 2003, many exotic candidates have been discovered.
With the gradual increase of the experimental statistics in the bottom hadron energy region, its three-body decay mode provides richer physics. For instance, its three-body decay modes can provide a chance to study charge asymmetry~\cite{BaBar:2005qms}, $CP$ asymmetry~\cite{BaBar:2006hyf,BaBar:2009vfr}, and the exotic states in the subsystems~\cite{BaBar:2009pnd,LHCb:2015klp,LHCb:2016axx,LHCb:2016nsl,LHCb:2020bls,LHCb:2020pxc,LHCb:2021uow,LHCb:2022aki,LHCb:2024vfz}.
These exotic states cannot be accepted by the traditional quark model, and their properties have been studied extensively. However, most of them focus on the two-body subsystem in the three-body decays of B mesons, e.g., Refs.~\cite{Braaten:2004ai,Dai:2015bcc,Li:2015iga,Duan:2023qsg,Ding:2025uhh}. In order to extract the resonance parameters with high precision, we should carefully deal with the final state interactions (FSI), especially the three-body FSI. A standard tool for three-body decays with FSI is Khuri-Treiman (KT) equations~\cite{Khuri:1960zz,Aitchison:1977ej}, which is proposed by Khuri and Treiman based on the analyticity and unitarity of the $S$-matrix. It has been successfully applied to the $K\to3\pi$ and $\omega\to3\pi$ as well as to extract the hadronic vacuum polarization in the anomalous
magnetic moment of the muon $(g-2)_\mu$~\cite{Holz:2024lom,Aoyama:2020ynm,Hoferichter:2018dmo}.

For a long time, it was only used for the interaction of kaons and pions~\cite{Gasser:2018qtg,Bernard:2024ioq,Alves:2024dpa,JPAC:2023nhq, Kou:2023kvp,DAmbrosio:2022jmd,Zhang:2021hcl,Niecknig:2017ylb}, because that the formalism needs their phase shift as input. 
It is achievable in experiment to perform a directly measurement of phase shift from two-body scattering process, due to the stability of pions and kaons. 
As pion and kaon only decay via weak decay, they can be considered as stable particles. The experimentalists first perform direct measurements of the $\pi\pi$ scattering to extract their phase shifts~\cite{Colangelo:2001df,Kaminski:2006qe}. 
 Therefore, the formalism was widely used in the final states involving $3\pi$. With the increasing experimental statistics, the phase shift of the $K\pi$ scattering is also extracted~\cite{Buettiker:2003pp},
 making it feasible to investigate the three-body FSI with kaon as one of final states~\cite{Niecknig:2015ija}. This approach can also be applied to the processes with the particles other than pions and kaons. For instance, recently, JAPC Collaboration analyze the BESIII data of the $e^+e^-\to J/\psi \pi^+\pi^-$ and $e^+ e^- \to \bar{D}D^*\pi$ processes. They extract the $Z_c(3900)$ resonance  parameters by fitting the $J/\psi\pi$ and $\bar{D}D^*$ invariant mass spectrum, i.e. taking into account the influence of the intermediate $D_0(2400)$ and $D_1(2420)$ mesons on the $\bar{D}D^*$ scattering, and the influence of the $f_0(980)$ and $\sigma$ on $J/\psi \pi$ scattering~\cite{Pilloni:2016obd}.

Recently, LHCb Collaboration observed two structures, $\chi_{c0}(3930)$ and $\chi_{c2}(3930)$~\cite{LHCb:2020pxc,LHCb:2020bls} in the invariant mass spectrum of $D^+D^-$ of the $B^+\to D^+D^-K^+$ process.
The mass, width and  quantum numbers of $\chi_{c0}(3930)$ are measured to be $M=3923.8\pm 1.5 \pm 0.4\  \mathrm{MeV}$, $\Gamma=17.4\pm5.1\pm0.8\ \mathrm{MeV}$ and $J^{PC}=0^{++}$. While those of the $\chi_{c2}(3930)$ are $M=3926.8\pm2.4\pm0.8\ \mathrm{MeV}$, $\Gamma=34.2\pm6.6\pm1.1\ \mathrm{MeV}$ and $J^{PC}=2^{++}$. They have been proposed as candidates for the potential tetraquark states~\cite{Wang:2020elp,Deng:2023mza,Ding:2024dif,Chen:2023eix,Shi:2021jyr,Duan:2020tsx,Ji:2022vdj}. As the positions of $\chi_{c0}(3930)$ and $\chi_{c2}(3930)$ located above the $D_s^+D_s^-$ threshold, they should be discovered in the invariant mass spectrum of $D_s^+D_s^-$ ~\cite{Prelovsek:2020eiw, Xin:2022bzt,Liu:2021xje,Mutuk:2022ckn}. Two years later, LHCb Collaboration indeed announces a structure $X(3960)$ in the $D_s^+D_s^-$ invariant mass distribution of the $B^{+}\to D_s^+D_s^-K^+$~\cite{LHCb:2022aki} process, but with mass at $M=3956\pm 5 \pm 10\ \mathrm{MeV}$. The $J^{PC}$ and width of the $X(3960)$ are $0^{++}$ and $\Gamma=43\pm 13 \pm8\ \mathrm{MeV}$ from the Breit-Wigner analysis. An interesting question is what the quantum number of $X(3960)$ is, either $\chi_{c0}(3930)$ or $\chi_{c2}(3930)$, and whether the two peak structures at $3.93~\mathrm{GeV}$ and $3.96~\mathrm{GeV}$ share the same dynamics.

As the phase space of the $B\to D\bar{D}K$ process is small, the FSI should be considered for precisely extracting resonances parameters, especially the three-body FSI.  
However, a direct extraction of the $D\bar{D}$, $\bar{D}K$ and $DK$ phase shifts in experiment is not feasible. For the past few years, people
have tried to extract the phase shift between heavy-flavor hadrons from lattice quantum chromodynamics(QCD)~\cite{Shi:2024llv}.
In addition, a combination of low-energy effective theory and lattice simulation also
calculate the phase shift of quasistable particle~\cite{Elhatisari:2021eyg}. In this work, we parametrize the $D\bar{D}$ scattering within the heavy quark spin symmetry (HQSS),  instead of calculating the $D\bar{D}$ phase shift  inserted into the Khuri-Treiman formalism, leaving the dynamic parameters  determined by experimental data. 

In this work, we precisely extract the resonance parameters of the $X(3960)$ or $\chi_{c0/2}(3930)$ with three-body FSI, 
by an overall fitting to the experimental data of the $B^+\to D^0\bar{D}^0K^+$, $B^+\to D^+D^-K^+$, and $B^+\to D_s^+D_s^-K^+$ processes, which indicates that they are the same state. 
 In Sec.~\ref{sec:Methods}, we introduce the kinematics and the dispersion representation of the three-body FSI, i.e. the KT formalism. The $D\bar{D}$ scattering amplitude is constructed within the heavy quark spin symmetry in Sec.~\ref{sec:Methods}. 
The fit results and discussions follow as Sec.~\ref{sec:Results}. The conclusion and outlook is presented in Sec.~\ref{sec:Summary}.

\section{Framework}
\label{sec:Methods}

\subsection{Kinematics}
\begin{figure*}[ht!]
\centering
    \subfigure[]{
    \includegraphics[width=0.30\textwidth]{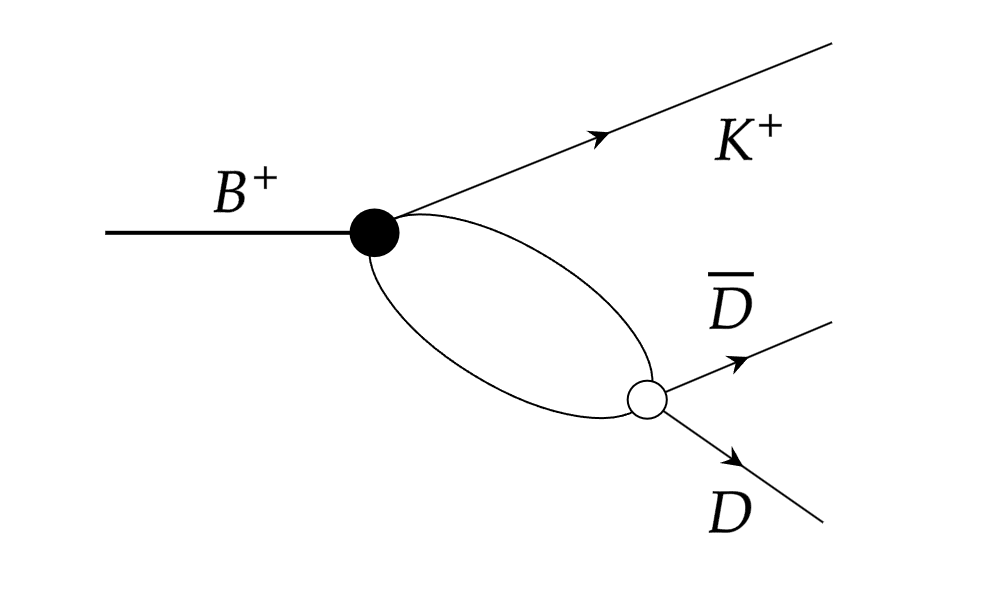}
    }
    \subfigure[]{
    \includegraphics[width=0.30\textwidth]{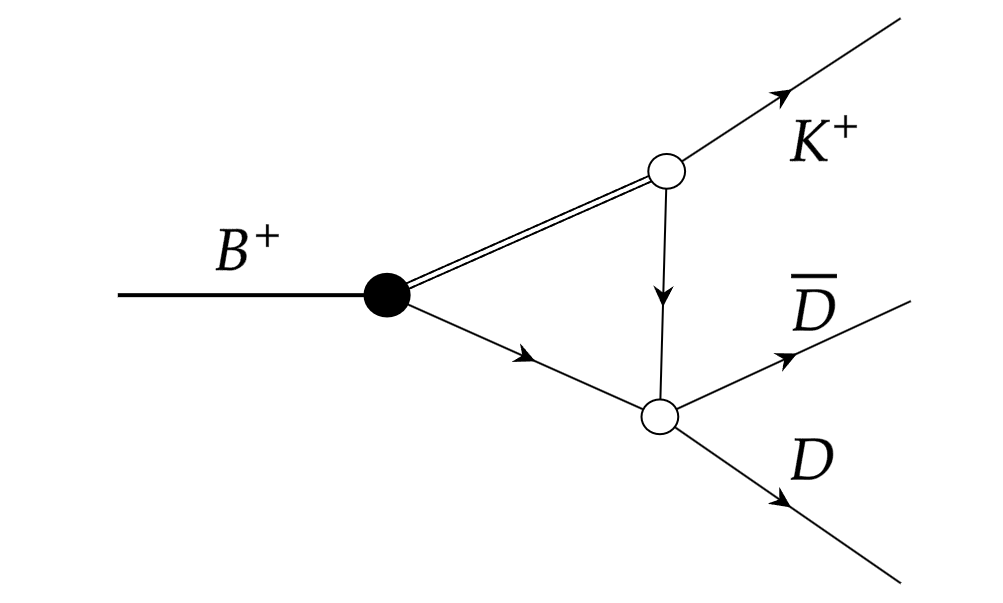}
    }
    \subfigure[]{
    \includegraphics[width=0.30\textwidth]{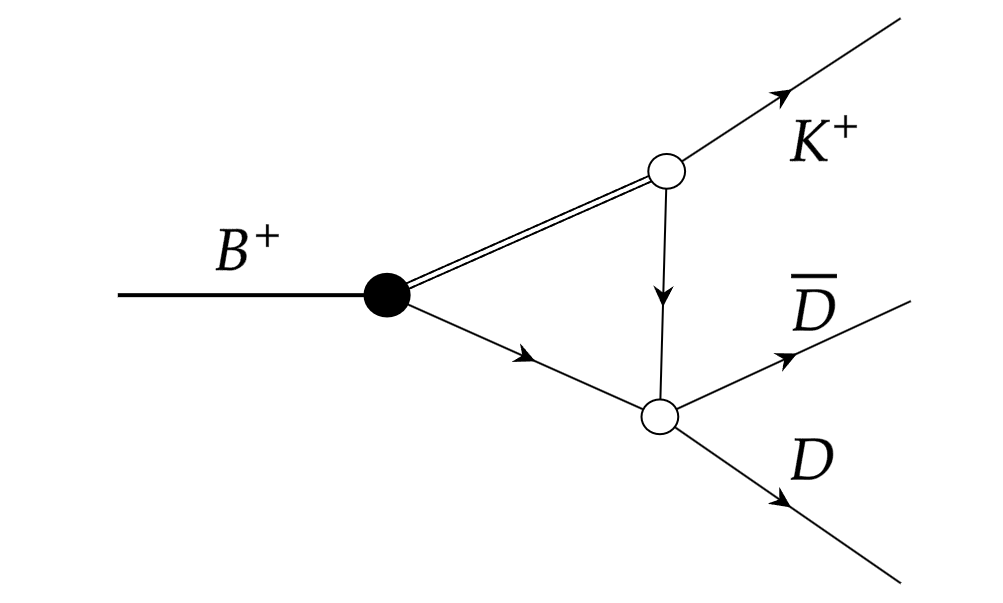}
    }
    \caption{The $B^+\to D\Bar{D}K^+$ decay reaction. (a) the contribution of $D\Bar{D}$ rescattering with right-hand cut; (b) the contribution of the resonance in the $\Bar{D}K^+$ channel projecting to $D\Bar{D}$ channel; (c) the contribution of the resonance in the $\Bar{D}K^+$ channel.}
    \label{fig:isobar}
\end{figure*}

In this section, before going into the details of KT formalism, the kinematic variables are defined for the further use. 
For the $B^{+}\to D\bar{D}K^{+}$ decays with $D=(D^0,D^+,D_s^+)$, 
there are two independent Lorentz invariant Mandelstam variables. Usually, three Mandelstam variables are defined as $s=(p_{D}+p_{\bar{D}})^2$, $t=(p_{\bar{D}}+p_K)^2$, and $u=(p_{D}+p_{K})^2$. The cosines of the helicity angles are defined as 
\begin{equation}
    \begin{aligned}
    &z_s=\frac{s(t-u)-(m_B^2-m_K^2)(m_{D}^2-m_{\bar{D}}^2)}{\lambda^{\frac{1}{2}}(s,m_{D}^2,m_{\bar{D}}^2)\lambda^{\frac{1}{2}}(s,m_{B}^2,m_{K}^2)},\\
    &z_t=\frac{t(s-u)-(m_B^2-m_D^2)(m_{K}^2-m_{\bar{D}}^2)}{\lambda^{\frac{1}{2}}(t,m_{K}^2,m_{\bar{D}}^2)\lambda^{\frac{1}{2}}(t,m_{B}^2,m_{D}^2)},\\
    &z_u=\frac{u(t-s)-(m_B^2-m_{\bar{D}}^2)(m_{D}^2-m_{K}^2)}{\lambda^{\frac{1}{2}}(u,m_{D}^2,m_{K}^2)\lambda^{\frac{1}{2}}(u,m_{B}^2,m_{\bar{D}}^2)},
\end{aligned}
\label{helicity angle}
\end{equation}
in the $D\bar{D}$, $\bar{D}K^+$, and $DK^+$ rest frames, 
with $\lambda(x^2,y^2,z^2)$ is kinematic K\"all\'en function,
\begin{align}
    \lambda(x^2,y^2,z^2)=(x^2-(y+z)^2)(x^2-(y-z)^2).
\end{align}
The decay amplitude $f_{i}(s,t,u)$ of the $i$th channel  can be decomposed into three isobar amplitudes
\begin{equation}
        f_i(s,t,u)=A^{(s)}_i(s,t,u)+A^{(t)}_i(s,t,u)+A^{(u)}_i(s,t,u),
\label{amplitude}
\end{equation}
with the $A_{i}^{(s,t,u)}$ the isobar amplitude of the $i$th decay channel, which describes the $D\bar{D}$, $\bar{D}K^+$, and $ DK^+ $ scattering as shown in Fig.~\ref{fig:isobar}. In our framework, we consider the resonance contribution of the $K^+\bar{D}$ channel. Furthermore, the $A_{i}^{(s)}(s,t,u)$ isobar amplitude contains the contributions from both  Fig.~\ref{fig:isobar}(a) and Fig.~\ref{fig:isobar}(b); The $A_{i}^{(t)}(s,t,u)$ isobar amplitude includes the resonance contribution in t-channel as shown by Fig.~\ref{fig:isobar}(c); As there is no resonance in u-channel, the contribution of the $A_{i}^{(u)}(s,t,u)$ isobar amplitude is neglected.
Here the channel index $i=1,2,3$ correspond to the $D^0\bar{D}^0$, $D^+D^-$ and $D_s^+D_s^-$ channels, respectively. 
Those can be expanded in terms of partial wave amplitudes 
\begin{equation}
    f_{i}(s,t,u)=16\pi\sum_{l=0}^{\infty}(2l+1)f_l^{i}(s)P_{l}(z_s),
\end{equation}
where $P_{l}(x)$ is the $l$th Legendre polynomials. The partial wave amplitudes can be decomposed into two parts, i.e. the right-hand cut part $a_{l,i}^{(s)}$ and the left-hand cut part $b_{l,i}^{(s)}$,
 \begin{align}
    f_{l}^{i}(s)=a_{l,i}^{(s)}(s)+b_{l,i}^{(s)}(s).
\end{align}
Here $a_{l,i}^{(s)}$ is the isobar partial wave amplitude 
\begin{equation}
    \begin{aligned}
        &a_{l,i}^{(s)}(s)=\frac{1}{32\pi}\int_{-1}^{1}dz_s\ A_{i}^{(s)}(s,t,u)P_{l}(z_s),
    \end{aligned}
    \label{eq:isobaramp-a}
\end{equation}
with the right-hand cut, which stems from the unitarity of the $S$-matrix, on the complex plane.
The second part 
\begin{equation}
    \begin{aligned}
        &b_{l,i}^{(s)}(s)=\sum_{m=t,u}\frac{2l^{\prime}+1}{2}\int_{-1}^{1}dz_s\ a_{l^{\prime},i}^{(m)}(m)P_{l^{\prime}}(z_m)P_{l}(z_s)
    \end{aligned}
    \label{eq:isobaramp-b}
\end{equation}
is the contribution of the $t$ channel and the $u$ channel isobar partial wave amplitude projected into the $s$ channel. In this sense, the $b_{l,i}^{(s)}(s)$ reflects the influence of the third particle $K^+$ on the $D\bar{D}$ scattering. From the above equation, one can see that the right-hand cuts of the $t$ channel and $u$ channel become the left-hand cut of $b_{l,i}^{(s)}$ on the $s$ plane after the projection. 

\subsection{Dispersive representation}
Through dispersion relation, one can obtain~\cite{Pilloni:2016obd}
\begin{equation}
    a_{l,i}^{(s)}(s)=\sum_{j}t_{ij}^{l}(s)\left(c_{l,j}+\frac{s}{\pi}\int_{s_{\mathrm{th}}^{j}}^{\infty}ds^{\prime}\frac{\rho_j(s^{\prime})b_{l,j}^{(s)}(s^{\prime})}{s^{\prime}(s^{\prime}-s)} \right),
    \label{eq:dispersion relation}
\end{equation}
where $t_{ij}^{l}(s)$ is the $l$-wave $D\bar{D}$ scattering amplitude  with $i,j=1,2,3$ the channel indexes. $s_{\mathrm{th}}^j$ is the threshold of the $j$th channel. 
The detailed discussion of $t_{ij}^l(s)$ can be found in the next section. 
 The $c_{l,j}$ is a polynomial subtraction, which is related to the behavior of the amplitude at large $s$. In our case, we perform a once subtraction, i.e. $c_{l,j}$ as a constant. $\rho_j(s)$ is the two-body phase space of the $j$th channel. 

\begin{figure}[htbp]
    \centering
    \includegraphics[width=0.48\textwidth]{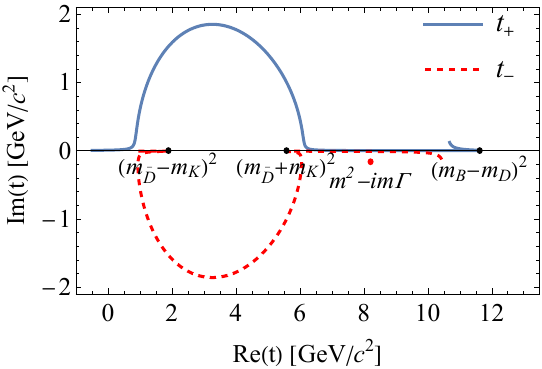}
    \caption{The dependence of $t_{\pm}$ on the variation of $s+i\epsilon$ on $t$-plane, with the arrow indicating the $s$ increasing direction. Here we assign a small positive imaginary part $\epsilon$ to $s$. The blue solid and red dashed curves are for $t_+$ and $t_-$, respectively. The red point is the singularity position of $a_{0,2}^{(t)}(t)$.}
    \label{tptm}
\end{figure}

\begin{figure*}[htbp]
    \centering
    \subfigure[The Breit-Wigner parametrization.]{
    \label{bl1s.sub.1}
    \includegraphics[width=0.95\textwidth]{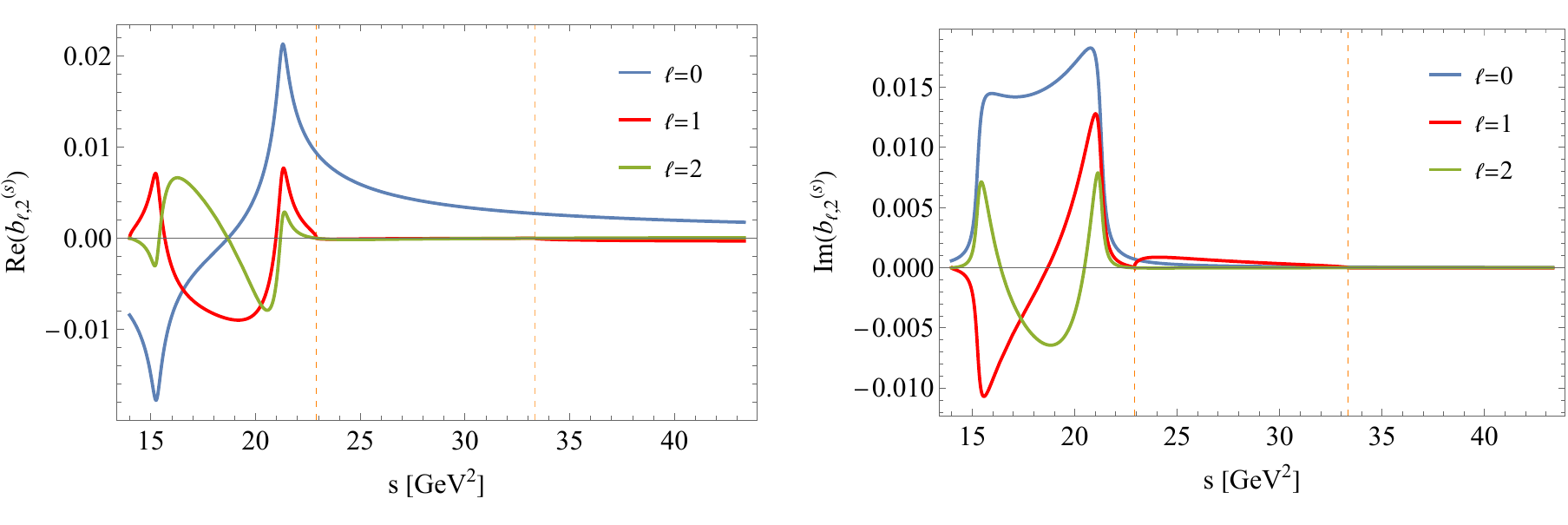}
    }
    \subfigure[The Flatt\'e parametrization.]{
    \label{bl1s.sub.2}
    \includegraphics[width=0.95\textwidth]{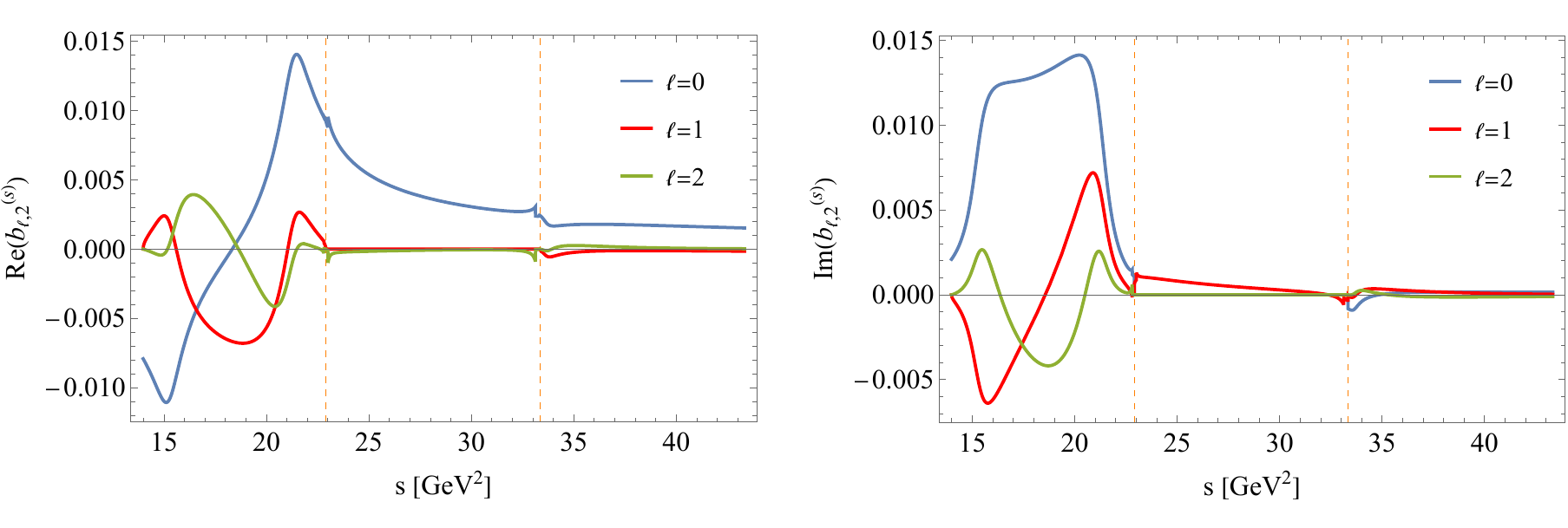}
    }
    \caption{The left and right pictures are  the real part and imaginary part of $b_{l,2}^{(s)}$, respectively. The blue, red and green curves are for $l=0,1,2$. The orange vertical dashed lines represent the positions of $(m_B-m_K)^2$ and $(m_B+m_K)^2$.}
    \label{bl1s}
\end{figure*}

From the experimental side, LHCb Collaboration observes a structure in the $D^-K^+$ invariant mass spectrum at about $2.9~\mathrm{GeV}$, which may be either isopin singlet spin-0 or isospin singlet spin-1 state~\cite{LHCb:2020pxc}. 
In this work, 
we include both cases in the $D^-K^+$ channel and ignore the potential resonance in the $D^+K^+$ channel, because the experimentalists do not find a peak structure in the $D^+K^+$ invariant mass spectrum. 
In this case, we only consider the influence of the $X_0(2900)$ or the $X_1(2900)$ in the $t$-channel and use a Breit-Wigner formula to parametrize its contribution~\cite{Pilloni:2016obd}, taking the $X_0(2900)$ as an example, 
\begin{equation}
    a_{0,2}^{(t)}(t)=\frac{1}{16\pi}\frac{g^2}{m^2-im\Gamma-t},
    \label{eq:BW-formula}
\end{equation}
where $m=2.893\  \mathrm{GeV}$, $\Gamma=0.057\ \mathrm{GeV}$, and $g$ are its  mass, width, and coupling to the $D^-K^+$ channel. Here $g$ is a free parameter. As currently, the signal of the $X_0(2900)$ or $X_1(2900)$ is reported in the $D^-K^+$ channel, we only include its contribution in this channel. As the result, only the 
$a_{0,2}^{(t)}(t)$ has a singularity on the t-plane. This Breit-Wigner formula would break unitary of $S$-matrix, when it has a close nearby channel or other overlap  resonances. However, this unitarity breaking effect is marginal in our case. 
We compare the two cases, i.e., the Breit-Wigner formula and Flatt\'e formula for the total width in Eq.~\eqref{eq:BW-formula}. 
In the former case, the total width is dealt as a constant. In the latter case, the width is considered saturated by the $D^-K^+$ channel, i.e.,  
\begin{equation}
    \Gamma(t)=\frac{1}{M}g^2\rho_{2}(t).
\end{equation}
One can see that the above Flatt\'e  amplitude 
\begin{equation}
    a_{0,2}^{(t)}(t)=\frac{1}{16\pi}\frac{g^2}{m^2-im\Gamma(t)-t},
\end{equation}
has a right-hand cut from the threshold to infinity, and the parametrization preserves the unitary of $S$-matrix. With the expression of $a_{l,i}^{(t)}(t)$, 
one can obtain the 
$b_{l,2}^{(s)}(s)$ from 
Eq.(\ref{eq:dispersion relation}) by integrating from $s_{th}$ to infinity.
It is a more convenient to calculate $b_{l,2}^{(s)}(s)$ for the integral of Eq.\eqref{eq:isobaramp-b} in the $t$-plane, because the singularity of $a_{0,2}^{(t)}$ is located in the $t$-plane. We  change the variable $z_s(dz_s)$ by the variable $t(dt)$. According Eq.~\eqref{helicity angle}, one can obtain 
\begin{equation}
    dz_{s}=\frac{2s}{\lambda^{\frac{1}{2}}(s,m_D^2,m_{\bar{D}}^2)\lambda^{\frac{1}{2}}(s,m_B^2,m_{K}^2)}dt. 
    \label{zs to t}
\end{equation}
The range $z_s\in [-1,+1]$ becomes $t\in [t_-,t_+]$. The $t_{\pm}$ depends on $s$ and is a complex number when $s\in [(m_{D}+m_{\bar{D}})^2, \infty]$. The presence of $t_{\pm}$ on the variable $s$ is as in Fig.~\ref{tptm}.
When $z_s$ goes from $-1$ to $1$, $t$ goes from $t_{-}$ to $t_{+}$. This process will not cross the singularity, so we do not need to change our integral path when we use the Breit-Wigners formula. If we use the Flatt\'e formula, the $t$ plane will appear as a cut line from $(m_{\bar{D}}+m_K)^2$ to infinity. To obtain a stable and correct   integration on the physical Riemann sheet, we need to change the integral path and make sure that the path does not cross the cut line. The comparison of $b_{l,2}^{(s)}(s)$ with these two parametrization schemes are presented in  Fig.~\ref{bl1s}.
The results within these two schemes behave similarly. The regions near the values of $(m_B-m_K)^2$ and $(m_B+m_K)^2$ in Fig.~\ref{bl1s.sub.2} exhibit strange behaviors because of their singularity behavior in the numerical calculation when the substitution, i.e., Eq.~\eqref{zs to t},  is performed. In this case, we use Breit-Wigner parametrization for our case.

 Using Eq.~\eqref{eq:isobaramp-a} and Eq.~\eqref{eq:dispersion relation}, we define
\begin{align}
    H_{l}(s)=\frac{s}{32\pi^2}\int_{s_{th}}^{\infty}ds^{\prime}\frac{\rho_{2}(s^{\prime})}{s^{\prime}(s^{\prime}-s)}\int_{-1}^{1} dz_s\frac{g^2P_l(z_s)}{m^2-im\Gamma-t}.
\end{align}
With the above definition, 
\begin{align}
    a_{l,i}^{(s)}(s)=t_{i1}^{l}(s)c_{l,1}+t_{i2}^{l}(s)[c_{l,2}+H_l(s)]+t_{i3}^{l}(s)c_{l,3}
\end{align}
and the expressions for the full decay amplitudes are given by
\begin{equation}
    \begin{aligned}
        &f_1(s,t,u)=16\pi [a_{0,1}^{(s)}(s)+3a_{1,1}^{(s)}(s)z_s],\\
        &f_2(s,t,u)=16\pi[a_{0,2}^{(s)}(s)+3a_{1,2}^{(s)}(s)z_s]+\frac{g^2}{m^2-im\Gamma-t},\\
        &f_3(s,t,u)=16\pi[a_{0,3}^{(s)}(s)+3a_{1,3}^{(s)}(s)z_s]
    \end{aligned}
    \label{eq:full amplitude}
\end{equation}

\subsection{$D\bar{D}$ interaction and Lippmann Schwinger equation}

In this section, we present the two-body interaction of the s-channel, i.e., the $D\bar{D}$ interaction, based on the heavy quark spin symmetry (HQSS). In our framework, both the contact potential and the normal charmonium contributions are considered. For the former one, 
the leading-order Lagrangian is given by~\cite{Ji:2022uie}
\begin{equation}
\begin{aligned}
    &\mathcal{L}_{4H} =-\frac14\,\mathrm{Tr}\left[H_{}^{a\dagger} H_{b}^{}\right]\mathrm{Tr}\left[\bar{H}_{}^{c} \bar{H}_{d}^{\dagger}\right] \left( F_{A}^{}\, \delta_a^{\,b} \delta_{c}^{\,d} + F_{A}^{\lambda}\, \vec\lambda_a^{\,b} \cdot \vec\lambda_c^{\,d} \right) \\
    &+\frac14\, \mathrm{Tr}\left[H_{}^{a\dagger} H_{b}^{} \sigma^{m}\right] \mathrm{Tr}\left[\bar{H}_{}^{c} \bar{H}_{d}^{\dagger} \sigma^{m}\right] \left( F_{B}^{}\, \delta_a^{\,b} \delta_{c}^{\,d} + F_{B}^{\lambda}\, \vec\lambda_a^{\,b} \cdot \vec\lambda_c^{\,d} \right),
\end{aligned}
\label{eq:la_ct}
\end{equation}
where the two-doublet superfields $H_a=P^{(Q)}_a+\sigma\cdot\vec{P}^{*(Q)}_a$ and $\bar{H}_a=P^{(\bar{Q})}_a+\sigma\cdot\vec{P}^{*(\bar{Q})}_a$ represent for charmed and anticharmed mesons. Here $a$ is the light quark flavor index. The $\sigma^i$ and $\lambda^i$ are the $i$th Pauli and Gell-mann matrices. $F_{A,B}^{(\lambda)}$ are the light-flavor-independent low-energy constants. The $P_a^{(Q)}$ and $P_a^{(\bar{Q})}$ are annihilation operators for charmed and anticharmed pseudoscalar mesons,
\begin{equation}
    P_a^{(Q)}=\begin{pmatrix}
        D^0\\
        D^{+}\\
        D_s^{+}
    \end{pmatrix},
    \quad\quad
    P_a^{(\bar{Q})}=\begin{pmatrix}
        \bar{D}^0\\
        D^{-}\\
        D_s^{-}
    \end{pmatrix}.
\end{equation}
The meaning of the $\vec{P}_a^{*(Q)}$ and $\vec{P}_a^{*(\bar{Q})}$ are analogous. 
From the above Lagrangian, one can extract the $S$-wave contact potential
\begin{equation}
    V^{S}_{\mathrm{CT}}=\begin{pmatrix}
        F_{A}+\frac{4}{3}F_{A}^{\lambda} & 2F_{A}^{\lambda} & 2F_{A}^{\lambda}\\
        2F_{A}^{\lambda} & F_{A}+\frac{4}{3}F_{A}^{\lambda} & 2F_{A}^{\lambda}\\
        2F_{A}^{\lambda} & 2F_{A}^{\lambda} & F_{A}+\frac{4}{3}F_{A}^{\lambda}
    \end{pmatrix}
\end{equation}
for the $D^0\bar{D}^0$, $D^+D^-$, $D_s^+D_s^-$ dynamic channels in the $\mathrm{SU}(3)_\mathrm{f}$ flavor symmetry. One can see that there are two free parameters $F_A$ and $F_A^\lambda$ in the potential. 
Once the $\mathrm{SU}(3)_\mathrm{f}$ flavor symmetry breaking effect is considered, the number of the parameters comes back to six. 
In this case, the $S$-wave  potential becomes 
\begin{equation}
    V^{S}_{\mathrm{CT}}=\begin{pmatrix}
        v_{11}^{0} & v_{12}^{0} & v_{13}^{0}\\
        v_{12}^{0} & v_{22}^{0} & v_{23}^{0}\\
        v_{13}^{0} & v_{23}^{0} & v_{33}^{0}
    \end{pmatrix}
\end{equation}
with the six parameters $v^0_{ij} (i\le j)$ determined from the experimental data. 
Analogously, the $P$-wave contact potential to the next-leading-order is momentum dependent,
\begin{equation}
    V^{P}_{\mathrm{CT}}=\begin{pmatrix}
        v_{11}^{1} & v_{12}^{1} & v_{13}^{1}\\
        v_{12}^{1} & v_{22}^{1} & v_{23}^{1}\\
        v_{13}^{1} & v_{23}^{1} & v_{33}^{1}
    \end{pmatrix}(\vec{p}_1-\vec{p}_2)(\vec{k}_1-\vec{k}_2),
\end{equation}
where $\vec{p}_1$($\vec{k}_1$), $\vec{p}_2$ ($\vec{k}_2$) are momenta of the incoming (outgoing) $D$ and $\bar{D}$, respectively. The six parameters $v^1_{ij} (i\le j)$ are also determined from the experimental data. 

Within the interested energy region, there are also several normal charmonia which can couple to a pair of open charmed mesons, which also contribute to the $D\bar{D}$ interactions. 
The leading order Lagrangian for the $S$-wave charmed meson pair, i.e., $D^{(*)}\bar{D}^{(*)}$ coupling to the $P$-wave charmonium is given as~\cite{Guo:2010ak}
\begin{align}
    \mathcal{L}_{\chi}=i\frac{g_0}{2}\left\langle \chi^{\dagger i}H_a \sigma^{i}\bar{H}_a\right\rangle+\mathrm{H.c.}.
\end{align}
Here, 
\begin{align}
    \chi^i=\sigma^j (-\chi_{c2}^{ij}-\frac{1}{\sqrt{2}}\epsilon^{ijk}\chi_{c1}^k+\frac{1}{\sqrt{3}}\chi_{c0})+h_c^i,
\end{align}
is $P$-wave charmonium super field in the heavy quark limit,
where $\chi_{c2}^{ij}$, $\chi_{c1}^{k}$, $\chi_{c0}$, and $h_c^i$ annihilate the $\chi_{c2}$, $\chi_{c1}$, $\chi_{c0}$, and $h_c$ states, respectively.
The coupling of $D^{(*)}\bar{D}^{(*)}$ to the $S$-wave charmonium reads as~\cite{Guo:2010ak}
\begin{align}
    \mathcal{L}_{\psi}=\frac{ig_1}{2}\left(\left\langle J^{\dagger}H_a\vec{\sigma}\cdot\partial \bar{H}_a\right\rangle-\left\langle J^{\dagger}\vec {\sigma}\cdot\partial H_a\bar{H}_a\right\rangle \right)+\mathrm{h.c},
\end{align}
where $J=\eta_c+\vec{\psi}\cdot\vec{\sigma}$ is the $S$-wave charmonium doublet in the heavy quark limit.  $\eta_c$ and $\vec{\psi}$ are the annihilation operators of the corresponding charmonia. 
In this case, the charmonium contributes to the $D\bar{D}$ interaction as 
\begin{equation}
\begin{aligned}
    &V_\mathrm{b}^{S}=\frac{g_0^2}{s-m_0^2-+im_0\Gamma_0},\\
    &V_\mathrm{b}^{P}=\frac{g_1^2(\vec{p}_1-\vec{p}_2)(\vec{k}_1-\vec{k}_2)}{s-m_1^2+im_1\Gamma_1},
\end{aligned}
\end{equation}
with the subindex $\mathrm{b}$ bare state. $m_l$ and $\Gamma_l$ are the mass and width of the bare state in $l$-wave $D\bar{D}$ interaction.
Because those charmonia are $\mathrm{SU}(3)_\mathrm{f}$ flavor symmetry singlets, they contribute to the three dynamical channels equally.

\begin{figure}[h]
    \centering
    \includegraphics[width=0.96\linewidth]{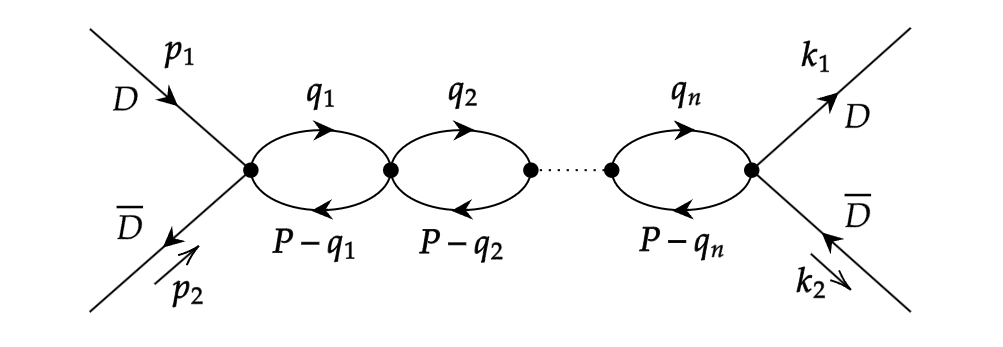}
    \caption{The n loop rescattering Feynman diagram for $D$ and $\bar{D}$. The black dots denote the potentials from both contact and bare state.}
    \label{fig:nloop}
\end{figure}

With the above potentials, the $D\bar{D}$ scattering amplitude can be expressed by sum over all the $n$-loop rescattering diagrams (as shown by Fig.~\ref{fig:nloop}) from $n=0$ to infinity, which procedure is encoded in Lippmann-Schwinger equation automatically.

The two-body propagator of the $l$-wave $D\bar{D}$ interaction for the $i$th channel reads as 
\begin{equation}
    G^{l}_i(s)=\int \frac{d^3q}{(2\pi)^3}\frac{|\mathbf{q}|^{2l}f^2_{\Lambda}(\mathbf{q}^2)}{\sqrt{s}-m_{D^i}-m_{\bar{D}^i}-\mathbf{q}^2/2\mu_i},
\end{equation}
where $\mu_i$ is the reduced mass of $i$th channel. $m_{D^i}$ and $m_{\bar{D}^i}$ are the masses of the $D$ and $\bar{D}$ mesons in the $i$th channel. Here, we use the nonrelativistic approximation, as we are interested in the near-threshold physics. 
The exponential form factor  
\begin{equation}
    f_{\Lambda}(\mathbf{q}^2)=\mathrm{exp}(-\mathbf{q}^2/\Lambda^2),
\end{equation}
is included to each vertex, as well as the vertex with the external particles.    
Here $\Lambda$ is a soft cutoff. 
For the lowest three partial waves, the two-body propagators read as
\begin{widetext}
\begin{align}
        G^{S}(s)&=-\frac{\mu\Lambda}{(2\pi)^{3/2}}+\frac{\mu k}{2\pi}e^{-2k^2/\Lambda^2}\left[ \mathrm{erfi} \left( \frac{\sqrt{2}k}{\Lambda}\right)-i\right],\\
        G^{\mathrm{P}}(s)
        &=-\frac{\mu\Lambda}{(2\pi)^{3/2}}\left(k^2+\frac{\Lambda^2}{4}\right)+\frac{\mu k^3}{2\pi}e^{-2k^2/\Lambda^2}\left[ \mathrm{erfi} \left( \frac{\sqrt{2}k}{\Lambda}\right)-i\right],\\
    G^{D}(s)&=-\frac{\mu\Lambda}{(2\pi)^{3/2}}\left(k^4+\frac{k^2\Lambda^2}{4}+\frac{3\Lambda^4}{16}\right)+\frac{\mu k^5}{2\pi}e^{-2k^2/\Lambda^2}\left[ \mathrm{erfi} \left( \frac{\sqrt{2}k}{\Lambda}\right)-i\right].
\end{align}
\end{widetext}
Finally, we can define the effective potential,
\begin{align}
    (V_{\mathrm{eff}}^{l})_{ij}=v^{l}_{ij}+\frac{g_{l,i}g_{l,j}}{s-m_l^2+im_l\Gamma_l},
\end{align}
where $i$, $j$ are the channel indexes. $m_{l}$, $\Gamma_{l}$ and $g_{l,i}$ are the mass, width and couplings of the bare state in the $l$-wave $D^{i}\bar{D}^{i}$ interaction,  respectively. 
 If the $\mathrm{SU}(2)_{f}$ symmetry holds pretty well, then $g_{l,1}=g_{l,2}=g_l$. A sizable $\mathrm{SU}(3)_\mathrm{f}$ breaking effect is reflected in the parameter  $r_1$ and $r_2$. For the $D^{i}\bar{D}^{i}\to D_{s}^{+}D_{s}^{-}$ scattering, the effective bare couplings are $g_{l,i}=: r_1 g_l$ for $i=1,2$ and $g_{3,3}=:r_2 g_l$ for $i=3$. The parameters $r_1$ and $r_2$ are used to describe the $\mathrm{SU(3)_\mathrm{f}}$ breaking effect.  
In this case, the $l$-wave $D\bar{D}$ scattering amplitude can be obtained as
\begin{align}
    t^{l}(s)=R^l (s)[1-V_{\mathrm{eff}}^{l}(s)G^{l}(s)]^{-1}V_{\mathrm{eff}}^{l}(s)R^l(s),
    \label{eq:t-matrix}
\end{align}
by solving Lippmann-Schwinger equation. Here $R^l$ is a factor matrix to guarantee the unitarity requirement of the $S$-matrix. As the LSE is in Lorentz scalar form, the momentum dependence
is encoded in the $R^l$ matrix. Analogously, 
the exponential form factor at each vertex should also be considered in the $R^l$ matrix to guarantee the unitarity requirement of the $S$-matrix. In this case, the matrix $R^l=\mathrm{diag}\{k_1^l e^{-(k_1^2/\Lambda^2)}, k_2^l e^{-(k_2^2/\Lambda^2)}, k_3^l e^{-(k_3^2/\Lambda^2)}\}$ with $k_i$ the three momentum for the $i$th channel in the $D\bar{D}$ rest frame.   

With those $D\bar{D}$ scattering amplitudes, one can obtain the full decay amplitude $f_i(s,t,u)$ from Eq.~\eqref{eq:full amplitude}  and 
the differential decay width as
\begin{align}
    \frac{d\Gamma_i}{d\sqrt{s}}=\frac{1}{(2\pi)^3}\frac{\sqrt{s}}{16m_B^3}\int_{t_-}^{t_+}dt|f_{i}(s,t,u(s,t))|^2,
\end{align}
where $t_{\pm}$ are the upper and lower limits of the Manderstanlm variable $t$. Using Eq.~\eqref{helicity angle} and $s+t+u=m_B^2+m_D^2+m_{\bar{D}}^2+m_K^2=:M^2$, $t_{\pm}$ can be expressed as
\begin{align}
    t_{\pm}=\frac{1}{2}\left(M^2-s\pm \frac{1}{s}\lambda^{1/2}(s,m_D^2,m_{\bar{D}}^2)\lambda^{1/2}(s,m_B^2,m_K^2)\right).
\end{align}
Analogously, for the t-channel, the corresponding $s_{\pm}$ is defined as
\begin{align}
    s_{\pm}=\frac{1}{2}\left(M^2-t\pm \frac{1}{t}\lambda^{1/2}(s,m_K^2,m_{\bar{D}}^2)\lambda^{1/2}(s,m_B^2,m_D^2)\right).
\end{align}
To compare with the experimental event distributions, we introduce three normalization constants, $a_i$ with $i = 1,2,3$ for the $i$th channel.  At the end, the event distributions read as 
\begin{align}
    \frac{dN_i}{dsdt}=a_i\frac{d\Gamma_i}{dsdt}.
    \label{events}
\end{align}
One notice that only two of those three normalization constants are independent parameters. 

\begin{figure*}[htp]
\centering
    \includegraphics[width=0.96\textwidth]{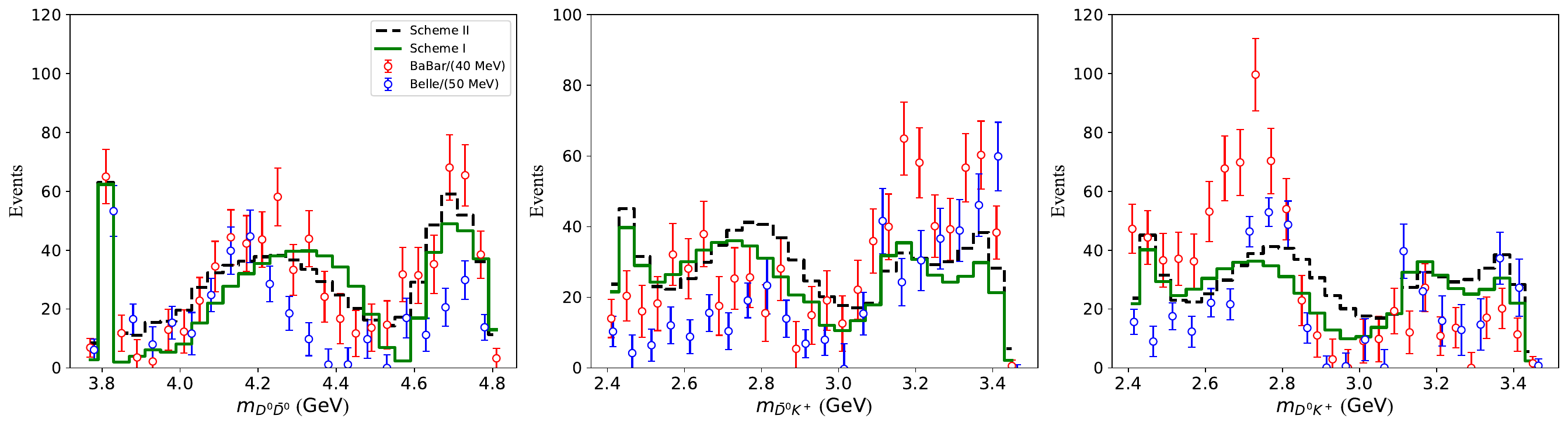}
    \includegraphics[width=0.96\textwidth]{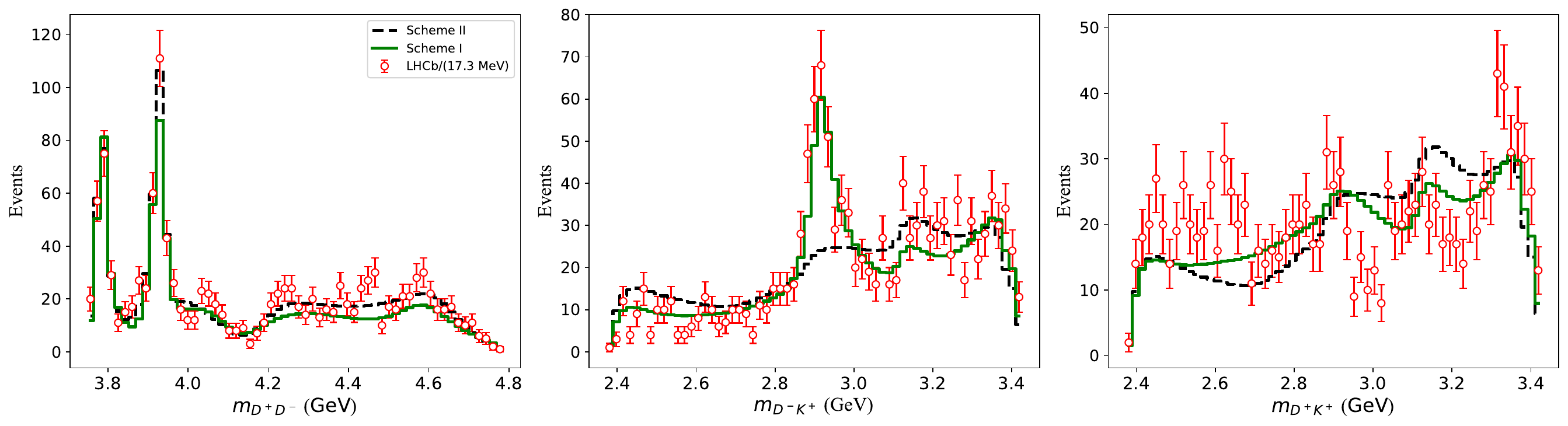}
    \includegraphics[width=0.96\textwidth]{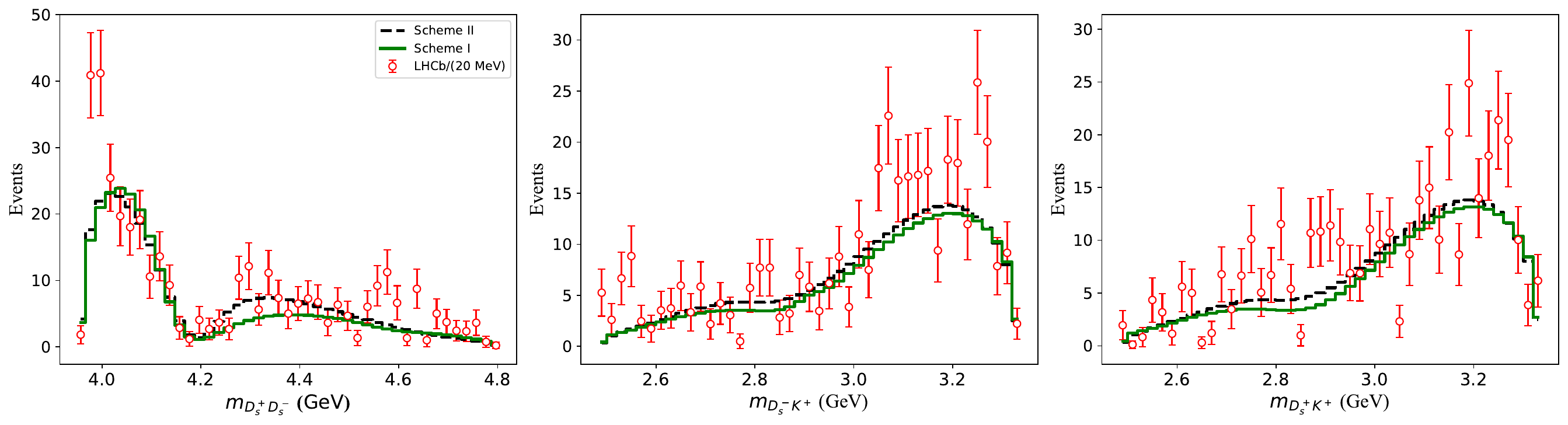}
    \caption{The invariant mass spectrum of $D\bar{D}$, $\bar{D}K$, and $DK$. The first line are the  fit result of $B^+\to D^0\bar{D}^0K^+$, where we plot the experiment result of $BABAR$ and Belle, which are represented in red and blue hollow point; The second line are the result of $B^+\to D^+D^-K^+$, the red hollow point come from LHCb experiment; The third line are the result of $B^+\to D_s^+D_s^-K^+$ channel, the red hollow point are experiment values come from LHCb experiment. And the black dashed and green solid step lines are the fit result of scheme II and scheme I, respectively.}
    \label{fig:fit result}
\end{figure*}

\section{Results and discussions}
\label{sec:Results}
\subsection{Fit results}
In this work, we fit the three processes of $B^{+} \to D\bar{D}K^+$ ($D\bar{D}$ channel means $D^0\bar{D}^0$, $D^+D^-$ and $D_s^+D_s^-$) simultaneously and perform a minimum $\chi^2/\mathrm{d.o.f}$ fit using MINUIT. Integration over $s$ or $t$ in Eq.~\eqref{events} is performed to obtain the $\bar{D}K^+$ or $D\bar{D}$ invariant mass spectra. 
We can also extract the $DK^+$ invariant mass distribution by choosing the independent kinematic variables as $s$ and $u$ and integrate over $s$. In this work, we choose $s$ and $t$ as the independent kinematic variables to extract the $D\bar{D}$ and $\bar{D}K^+$ invariant mass distributions. One can also use $u$ instead of $t$ to derive $DK^+$ invariant mass spectrum. The numerical integration is then performed through a 48-point Gauss-Legendre quadrature to accelerate it,
\begin{equation}
    \frac{dN_i}{d\sqrt{s}}(\sqrt{s})=a_i\frac{d\Gamma_i}{d\sqrt{s}}\to \frac{a_i}{(2\pi)^3}\frac{\sqrt{s}}{16m_B^3}\sum_{n=1}^{N}|f_i(s,t_n)|^2\omega_n,
\end{equation}
where $\omega_n$ are quadrature weights, and $N=48$,
\begin{equation}
    \begin{aligned}
        &\omega_n=\frac{b-a}{(N+1)^2}\frac{1-x_n^2}{P_{N+1}^2(x_n)},\\
        &P_{N}(x_n)\equiv 0,\\
        &t_n=\frac{t_+-t_-}{2}x_n+\frac{t_+ + t_-}{2}.
    \end{aligned}
\end{equation}
We fit the experimental data bin by bin. In practice, each bin is divided into ten small bins. In this case, the integration over the variables $s$ or $t$ is changed to sum over these ten small bins. 
\begin{equation}
\begin{aligned}
        &N_i(n)=\sum_{m=1}^5\frac{dN_i}{d\sqrt{s}}(E_m)\frac{w_i}{10},\\
        &E_m = \sqrt{s}_n+\frac{\sqrt{s}_{n+1}-\sqrt{s}_n}{10}(m-1),
\end{aligned}
\end{equation}
where $w_i$ is the weight of the $i$th bin, and $\sqrt{s}_n$ is energy of the $n$th bin.
The fit results are presented in Fig~.\ref{fig:fit result}.

We consider two fit schemes: scheme I, we consider the influence of either $X_{0}(2900)$ or $X_1(2900)$ using Eq.~\eqref{eq:full amplitude} to calculate the three-body final state interaction(FSI); scheme II, we ignore the influence of both $X_{0}(2900)$ and $X_1(2900)$, but consider only the $D\bar{D}$ two-body FSI, and only fit to the experimental $D\bar{D}$ invariant  mass distributions. The experimental data of the $B^+\to D^0\bar{D}^0K^+$ channel come from $BABAR$ and Belle Collaborations\cite{BaBar:2014jjr,Belle:2007hht}, and $B^{+}\to D_{(s)}^+D_{(s)}^-K^+$ channel date come from LHCb Collaborations\cite{LHCb:2020pxc,LHCb:2022aki}. However, we only use the data from $BABAR$ Collaboration for the $B^+\to D^0\bar{D}^0K^+$ fit due to its simpler background.
The initial values of the parameters of scheme I are the fit results of scheme II.  In this  cases, the amplitude reads as 
\begin{align}
    f_i(s,t,u)=16\pi\sum_{l,j}(2l+1)t_{ij}^{l}(s)c_{l,j}P_{l}(z_s),
\end{align}
where $j=1,2,3$ is the channel index, and $l=0,1$, is orbital angular momentum. The initial parameters of scheme II are chosen randomly. The final fit result of scheme II is shown as the black step line in Fig.~\ref{fig:fit result}, with $\chi^2/\mathrm{d.o.f}=1.758$. Since we only use the data of the $D\bar{D}$ channel for the fitting in scheme II, the corresponding $\chi^2$ is smaller than that of scheme I. The fitted parameters of scheme II are used as the initial values for scheme I to fit the  $D\bar{D}$, $\bar{D}K^+$ and $DK^+$ invariant mass distributions, and the obtained $\chi^2/\mathrm{d.o.f}=2.488$, $\Lambda=1.591~\mathrm{GeV}$, with the other parameters are listed in Appendix~\ref{app:fit parameters}. We also check the stability of our results by varying $\Lambda=1.432~\mathrm{GeV},~1.591~\mathrm{GeV},~1.750~\mathrm{GeV},~1.909~\mathrm{GeV},~2.068~\mathrm{GeV}$. The reduced chi squares corresponding to these five $\Lambda$ values are almost the same, which indicate that they would give similar line shapes, as well as other physical quantities. The details can be found in Appendix~\ref{app:B}. For marching the peak about $\sqrt{t}=2.9\ \mathrm{GeV}$ in the $D^-K^+$ channel, we assume a state of $J^{P}=0^+$, $m=2.893\ \mathrm{GeV}$, and $\Gamma=0.057\ \mathrm{GeV}$, but its influence on the $D\bar{D}$ scattering is very small.

The fit results indicate that
two bare states, i.e. $X_0(2900)$ in $t$-channel and $\psi(3770)$ in $s$-channel, are sufficient to explain the experimental data very well. Their masses, widths and couplings to the $D^0D^0$, $D^+D^-$ channels are presented in Table.~\ref{tab:bare state1} and Table.~\ref{tab:bare state2} for scheme I and scheme II, respectively. 
The couplings to the $D_s^+D_s^-$ channel 
are defined by the other parameters $r_1=0.462\pm0.009$ and $r_2=0.235\pm0.011$ as discussed in the previous section. Their values are obtained from the fit to the data for scheme I. Those are $r_1=0.354\pm0.012$, $r_2=0.200\pm0.053$ for scheme II.
We can extract  scattering length and effective range from effective-range-expansion (ERE)
\begin{eqnarray}
    t^{0}(s)=-\frac{2\pi}{\mu}\frac{1}{-1/a_0+1/2r_0k^2-ik+\mathcal{O}(k^4)},
\end{eqnarray}
where $s=(m_1+m_2+\frac{k^2}{2\mu})^2$, and $\mu=\frac{m_1m_2}{m_1+m_2}$ is reduced mass of $m_1$ and $m_2$. $a_0$ and $r_0$ are scattering length and effective range, respectively, defined at the lowest threshold. The superscript $0$ of $t^0(s)$ and subscripts $0$ of $a_0$, $r_0$ means the relative orbital angular momentum $L=0$, i.e., $S$-wave. One can extract them by   
\begin{equation}
    \begin{aligned}
        &a_0=\left. \frac{\mu}{2\pi}t^{0}(s)\right|_{s\to s_{th}},\\
        &r_0=-\left. \frac{2\pi}{\mu}\mathrm{Re}\left[\frac{d(t^{0}(\sqrt{s}))^{-1}}{d\sqrt{s}}\right]\right|_{\sqrt{s}\to\sqrt{s_{th}}},\\
        &r^{\prime}=r_0-\Delta r,
    \end{aligned}
\end{equation}
where $\Delta r$ is the couple channel corrections, and one should remove those corrections to obtain the compositeness of a given particle. How to extract this correction has been demonstrated  in Appendix~\ref{app: C}.
Furthermore, the Weinberg criterion, which is typically used to characterize the compositeness of bound, virtual and resonance states, can be found~\cite{Baru:2021ldu,Niu:2024cfn,Matuschek:2020gqe,Li:2021cue} as,
\begin{eqnarray}
    \bar{X}=\frac{1}{\sqrt{1+2|\frac{r^{\prime}}{\mathrm{Re}[a_0]}|}}.
\end{eqnarray}
According to the fit results, the scattering length, effective range and compositeness of the $S$-wave are $a_0=0.8355 -0.0004i$, $r_0=-6.2240$, $r^{\prime}=3.303$, and $\bar{X}=0.3545$. 
The compositeness indicates that $\chi_{c0}(3930)$ contains about $34\%$ molecular component, making it more like a compact normal charmonium.

\begin{table}[h!]
    \caption{The mass, width, and coupling of the bare state in scheme I (fitted values), the coupling represent the couple between the bare state with $D^0\bar{D}^0$ or $D^+D^-$.}
    \centering
    \renewcommand{\arraystretch}{1.5}
    \resizebox{0.48\textwidth}{!}{
    \begin{tabular}{|c|c|c|c|}
    \hline
         & Mass\ (GeV) & Width\ (GeV)& Coupling \\
         \hline
        l=0 &$3.909\pm0.002$ &$0.035\pm 0.006$& $12.409\pm 0.537$ \\
        l=1&$3.802\pm 0.006$ & $0.013\pm 0.004$& $17.465\pm 1.663$\\
        \hline
    \end{tabular}}
    \renewcommand{\arraystretch}{\origArraystretch}
    \label{tab:bare state1}
\end{table}

\begin{table}[h!]
    \caption{The mass, width, and coupling of the bare state (fitted values) in scheme II, the coupling represent the couple between the bare state with $D^0\bar{D}^0$ or $D^+D^-$.}
    \centering
    \renewcommand{\arraystretch}{1.5}
    \resizebox{0.48\textwidth}{!}{
    \begin{tabular}{|c|c|c|c|}
    \hline
         & Mass\ (GeV) & Width\ (GeV)& Coupling \\
         \hline
        l=0 & $3.892\pm0.003$& $0.029\pm0.005$&$22.157\pm0.624$\\
        l=1& $3.803\pm0.006$& $0.019\pm0.005$&$30.930\pm1.656$\\
        \hline
    \end{tabular}
    }
    \renewcommand{\arraystretch}{\origArraystretch}
    \label{tab:bare state2}
\end{table}

The mass positions are very close to the peaks of $X(3960)$ and $\psi(3770)$ in the $D\bar{D}$ channel, which stem from the input bare states discussed in the next section. 
The dip structures in the $D^+D^-$ and $D_s^+D_s^-$ channels around $\sqrt{s}=4.16~\mathrm{GeV}$,and the very broad peak in the $D^0\bar{D}^0$ channel stem from a nearby pole, which will be discussed in the next section. 

\subsection{Poles in the $D\bar{D}$ system}

With the parameters fitted from the experimental data in the last section, 
we perform the pole analysis of the $D\bar{D}$ system in this section. 
The poles of the $D\bar{D}$ system  
can be obtained by the zeros of the denominator of the $D\bar{D}$ scattering amplitude, i.e. Eq.~\eqref{eq:t-matrix}. 
The scattering amplitude can be rearranged as 
\begin{eqnarray}\nonumber
    t^{l}(s)&=&R^l (s)[1-V_{\mathrm{eff}}^{l}(s)G^{l}(s)]^{-1}V_{\mathrm{eff}}^{l}(s)R^l(s)\\\nonumber
    &=& \frac{1}{R^{l,-1}(s)[1-V^l_\mathrm{eff}(s)G(s)]R^l(s)} R^l(s)V^l_\mathrm{eff}(s)R^l(s),
    \label{eq:t-matrixV2}
\end{eqnarray}
from which one can see that the pole positions are determined by
\begin{align}
    \mathrm{Det}[1-V_{\mathrm{eff}}^l(s)G^{l}(s)]=0
    \label{eq:pole}
\end{align}
without corrections from the $R^l(s)$ factor matrix. We have three dynamical channels, i.e. $D^0\bar{D}^0$, $D^+D^-$, $D_s^+D_s^-$ channels, which are ordered by their thresholds from lower to higher one. 
In total, there are $2^3=8$ Riemann sheets (RSs) $\mathrm{R}_{\pm\pm\pm}$ accordingly, with the physical sheet $\mathrm{RS}_+$ and unphysical sheet $\mathrm{RS}_-$ defined by their three momentum.   
\begin{equation}
\begin{aligned}
        &\mathrm{RS}_+: k_i=\sqrt{2\mu_i(\sqrt{s}-m_{D^i}-m_{\bar{D}^i})},\\
    &\mathrm{RS}_-: k_i=-\sqrt{2\mu_i(\sqrt{s}-m_{D^i}-m_{\bar{D}^i})},
\end{aligned}
\end{equation}
with $m_{D^i}$, $m_{\bar{D}^i}$ the masses of the $D$ and $\bar{D}$ meson in the $i$th channel. $\mu_i$ is the reduced mass of the $i$th channel. 
We search for the poles on all the RSs and collect them in 
Appendix~\ref{app:z fun}. Here we only present the poles on the physical sheet $\mathrm{R}_{+++}$ and those close to the physical sheet in Table.~\ref{tab:pole_min}. 
The pole $4.106-0.092$ located on $\mathrm{RS}_{---}$, correspond to a resonant state. Additionally, it has a corresponding poles on the upper half-plane due to Schwarz reflection. However, we do not present the pole in Table.~\ref{tab:pole_min}, because they are far from the physical axis.
The pole at $3.913-0.016i$ is very close to the experimentally observed state $\chi_{c0}(3930)$ with $J^{PC}=0^{++}$. One can see that, in Table.~\ref{tab:mass&width for chi&psi}, the mass and width of the $\chi_{c0}(3930)$ are significantly channel-dependent, which is a very natural property of Breit-Wigner analysis. 
The pole $3.913-0.016i$ is consistent with the states reported in the $B^+\to D^+D^-K^+$ and $B^+\to D_s^+D_s^-K^+$ processes as expected. We have tried another potential quantum number, i.e. $J^{PC}=2^{++}$, but fail to obtain a good description of the experimental data. The pole in the $P$-wave scattering amplitude at $3.761-0.006i$ is close to the physical mass of $\psi(3770)$. The mass is consistent with the experimental measurement and the width is smaller than the experimental measurement, as shown in Table.~\ref{tab:mass&width for chi&psi}. The pole at $4.106-0.092i$ does not have significant physical contributions due to its large imaginary parts.
Furthermore, the below-threshold pole $3.5912 - 0.010i$ is far away from the lowest threshold and does not have influence on physical observables. 

\begin{table}[t]
    \caption{The masses and widths of $\chi_{c0}(3930)$ and $\psi(3770)$ measured in various reactions.}
    \centering
    \resizebox{0.48\textwidth}{!}{
    \begin{tabular}{|c|c|c|c|}
    \hline
         &Channel& Mass\ (MeV) & Width\ (MeV) \\
         \hline
        \multirow{7}{*}{$\chi_{c0}(3930)$}
        &$B^{+}\to D^+D^-K^+$\cite{LHCb:2020pxc}& $3923.8\pm1.5 \pm0.4$& $17.4\pm5.1\pm0.8$\\
        &$B^+\to D_s^{+}D_s^-K^{+}$\cite{LHCb:2022aki}&$3956\pm5\pm10$ &$43\pm13\pm 8$ \\
        &$B\to K\omega J/\psi$\cite{Belle:2004lle}& $3943\pm11\pm
 13 $&$87\pm22\pm26$\\
        &$e^+e^-\to J/\psi D\bar{D}$\cite{Belle:2017egg} &$3862^{+26+40}_{-32-13}$ &$201^{+154+88}_{-67-82}$\\

        &$\gamma\gamma\to\omega J/\psi$\cite{Belle:2009and}& $3915\pm3\pm2$&$17\pm10\pm3$\\
        &$\gamma\gamma\to \omega J/\psi$\cite{BaBar:2012nxg}&$3919.4\pm2.2\pm1.6$&$13\pm6\pm3$\\
        &$e^+e^-\to \gamma \omega J/\psi$\cite{BESIII:2019qvy}& $3932.6\pm8.7$&$59.7\pm15.5$\\
        \hline
       \multirow{2}{*}{$\psi(3770)$} 
       &$e^+e^-\to D\bar{D}$\cite{Anashin:2011kq} & $3779.2^{+1.8+0.5+0.3}_{-1.7-0.7-0.3}$&$24.9^{+4.6+0.5+0.2}_{-4.0-0.6-0.9}$\\
       &$B^+\to D^0\bar{D}^0K^+$\cite{Belle:2007hht} &$3776\pm5\pm4$ &$27 \pm10\pm5$\\
        \hline
    \end{tabular}
    }
    \label{tab:mass&width for chi&psi}
\end{table}
 Whether these poles are the renormalized bare poles or the dynamic generated poles can be easily seen by their trajectory on the $z$-plane, which will be discussed afterward.

\begingroup
\setlength{\extrarowheight}{4pt}
\begin{table}[b]
    \centering
    \renewcommand{\arraystretch}{1.3}
    \caption{The poles on various RSs. The real part is the energy and the imaginary part is the half width. The errors originate from the uncertainties of the cutoff and the parameters in the potential ~\footnote{The errors are obtained from boostrap method by generating 10000 samples.}. }
    \resizebox{0.45\textwidth}{!}{
    \begin{tabular}{|c|c|c|}
    \hline
        RS & $l=0$&$l=1$ \\
        \hline
       $+++$  & $3.525^{+0.144}_{-0.196}-0.007^{+0.005}_{-0.005}i$& \\
         \hline
      $--+$  & $3.913^{+0.003}_{-0.003}-0.016^{+0.003}_{-0.003}i$ &$3.761^{+0.031}_{-0.061}-0.006^{+0.006}_{-0.006}i$\\
      \hline
      $---$  & $4.106^{+0.066}_{-0.062}-0.092^{+0.037}_{-0.049}i$ &  \\
      \hline
    \end{tabular}
    }
    \renewcommand{\arraystretch}{\origArraystretch}
    \label{tab:pole_min}
\end{table}

In our framework, we introduce bare poles to the $D\bar{D}$ potential, which raises a question whether the above poles are dynamically generated state or from the input bare state.
One can figure this out by reproducing the trajectory of the renormalization, i.e., setting the bare couplings from the fitted values to zero. It is easy to understand from the one-channel case, but without losing generality. The effective potential $V_\mathrm{eff}^l$ for a given particle wave $l$ gains contributions from two parts, i.e., the contact potential part $V^l_\mathrm{CT}$ and the bare state part $\frac{g_l^2}{s-m_l^2+im_l\Gamma_l}$.  The scattering amplitude can be obtained by solving LSE as
\begin{widetext}
\begin{equation}
    \begin{aligned}
    t^l&\propto [1-(\frac{g_l^2}{s-m_l^2+im_l\Gamma_l}+V_\mathrm{CT}^l)G^l]^{-1}\\
    &=(1-V_\mathrm{CT}G^l)^{-1} \left( 1- (1-V_\mathrm{CT}G^l)^{-1}\frac{g_l^2}{s-m_l^2+im_l\Gamma_l} G^l\right)^{-1}\\
    &\propto (1-V_\mathrm{CT}G^l)^{-1} \left( (s-m_l^2+im_l\Gamma_l)- (1-V_\mathrm{CT}G^l)^{-1}g_l^2G^l\right)^{-1}.
    \end{aligned}
        \label{eq:poles}
\end{equation}
\end{widetext}
From the above equation, one can easily see that two kinds of poles contribute. The first kind of pole is only from the contact potential and determined by 
\begin{eqnarray}
1-V_\mathrm{CT}G^l=0.
\end{eqnarray}
The second kind of pole is from the renormalized bare pole contribution, which is determined by the second factor of Eq.~\eqref{eq:poles}, i.e. 
\begin{eqnarray}
    (s-m_l^2+im_l\Gamma_l)- (1-V_\mathrm{CT}G^l)^{-1}g_l^2G^l=0.
\end{eqnarray}
When the coupling $g_l$ is set to zero, the second kind of poles come back to the bare pole and determined by 
\begin{align}
s-m_l^2+im_l\Gamma_l=0.
\end{align}
In this case, when we decrease the coupling from given values to zeros, the second kind of poles will move to the bare pole positions and the pole trajectory reflect the renormalization.  
As the $S$-matrix is analytic in the whole complex energy plane on top of cuts and poles, the poles should move smoothly with the variation of a real parameter. It could move from one sheet to another one, but not disappear. To accurately present the trajectory, it would be very convenient to map the eight RSs on the complex $\sqrt{s}$ plane onto a uniformization variable $z$ for three channels following Ref.~\cite{Yamada:2022xam}. Before mapping, we define the ``momentum" $q_i\equiv \sqrt{s-\epsilon_{i}^{2}}$ of the $i$th channel with $\epsilon_i\equiv m_{D^i}+m_{\bar{D}^i}$, which ignores the pseudothreshold of the three momentum. As the masses of the two particles in the $i$th channel are equal to each other, the pseudothreshold is at $\sqrt{s}=0 ~\mathrm{GeV}$, making its contribution marginal. The kinematic parameter 
$\gamma\equiv\frac{\sqrt{\epsilon_{3}^{2}-\epsilon_{1}^{2}}+\sqrt{\epsilon_{3}^{2}-\epsilon_{2}^{2}}}{\sqrt{\epsilon_{2}^{2}-\epsilon_{1}^{2}}}$ is defined for the later use. $q_1$, $q_2$,  and $q_3$ are expressed as a single-valued functions of $z$ via Jacobi elliptic functions
\begin{equation}
q_1=\frac{\sqrt{\varepsilon_2^2-\varepsilon_1^2}}{2}[\frac{\gamma}{SN(4K(\frac{1}{\gamma^2})z,\frac{1}{\gamma^2})}+\frac{SN(4K(\frac{1}{\gamma^2})z,\frac{1}{\gamma^2})}{\gamma}],
\end{equation}
\begin{equation}
    q_2=\frac{\sqrt{\varepsilon_2^2-\varepsilon_1^2}}{2}[\frac{\gamma}{SN(4K(\frac{1}{\gamma^2})z,\frac{1}{\gamma^2})}-\frac{SN(4K(\frac{1}{\gamma^2})z,\frac{1}{\gamma^2})}{\gamma}],
\end{equation}
\begin{figure*}[htp]
    \centering
     \subfigure[The poles of S wave.]{
    \includegraphics[width=0.48\textwidth]{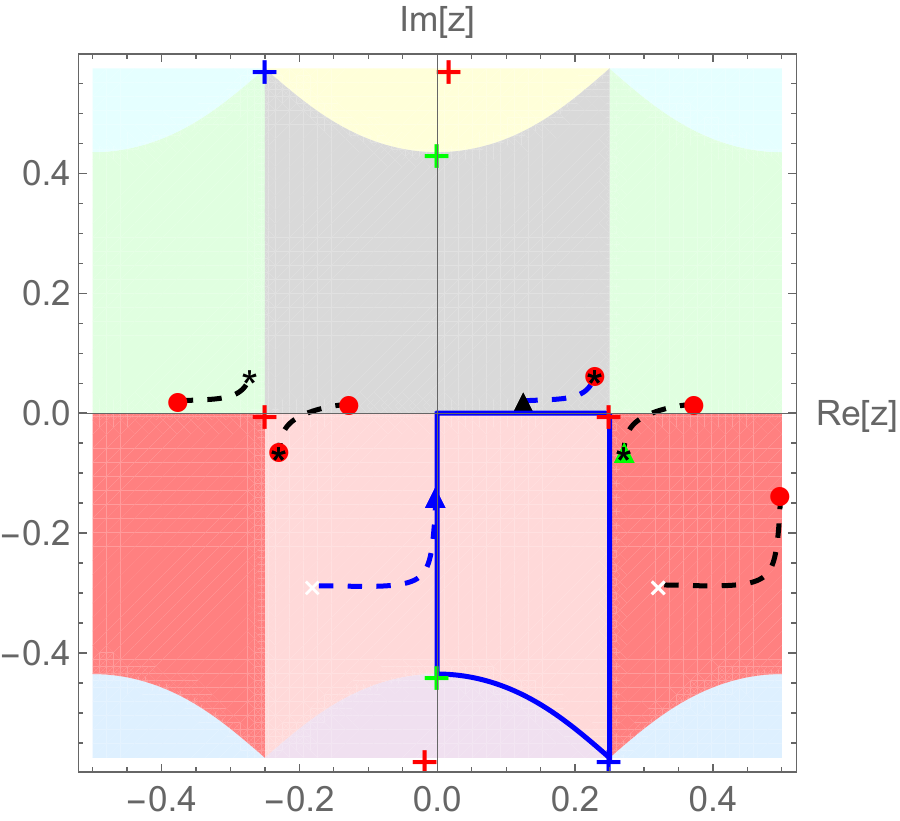}
    \label{subfig:pole_s}
    }
    \subfigure[The poles of P wave.]{
    \includegraphics[width=0.48\textwidth]{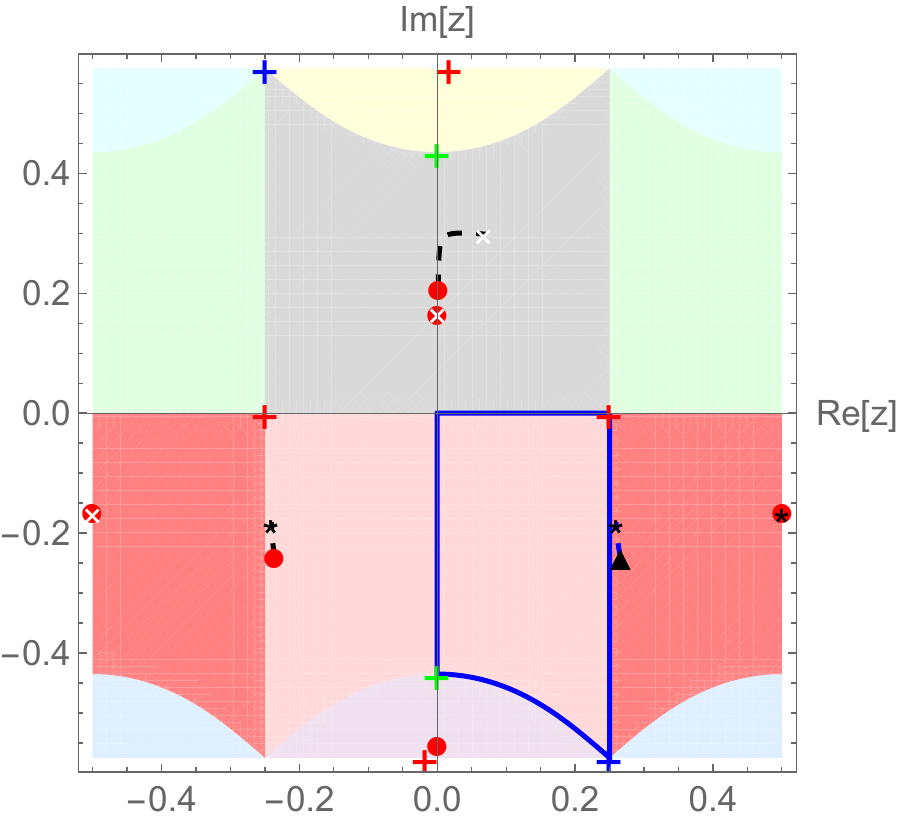}
    \label{subfig:pole_p}
    }
    \includegraphics[width=0.32\textwidth]{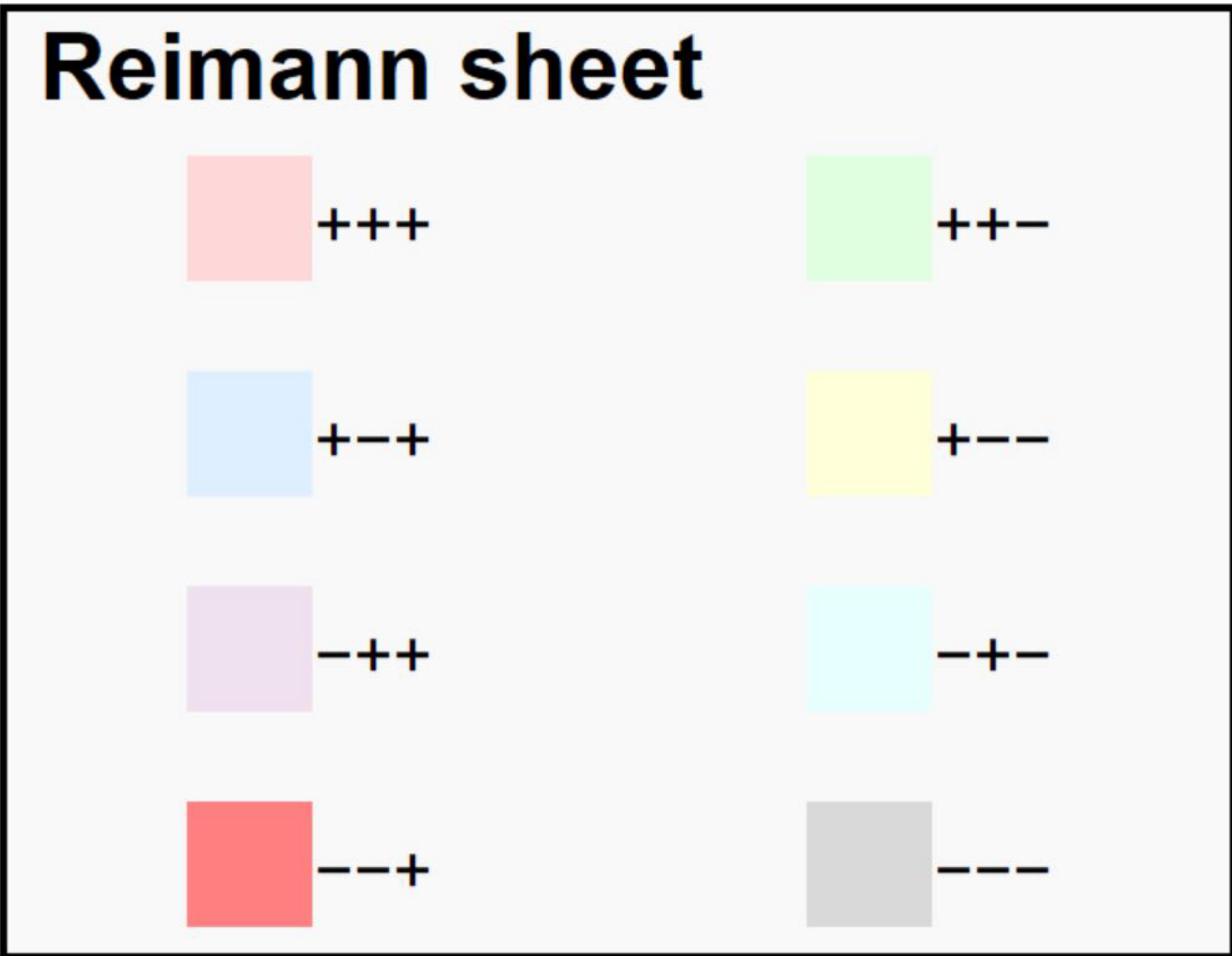}
    \caption{The conformal mapping from the complex $s$-plane to the complex $z$-plane. The eight RSs on $s$-plane are labeled by eight different colors. The left and right figures are for the $S$-wave and $P$-wave $D\bar{D}$ interactions, respectively. The blue solid lines are the physical axis where the physical  observables locate. The green, blue and red plus signs are the positions of the first, second and third thresholds, respectively. The red dots denote the poles far away from the physical region. The green triangles denote the poles close to the physical region. The black star dots represent the poles trending to the bare states when the bare state couplings tend to zero. The white crosses represent the poles did not return to the bare state. The blue dashed lines denote the trajectory of the poles close to the physical region, when the couplings increase. The black dashed lines are analogous to that of blue dashed ones, but for the poles far away from the physical region. }
    \label{fig:pole position}
\end{figure*}
\begin{equation}
    q_3=\frac{\sqrt{\varepsilon_2^2-\varepsilon_1^2}}{2}\frac{\gamma CN(4K(\frac{1}{\gamma^2})z,\frac{1}{\gamma^2})DN(4K(\frac{1}{\gamma^2})z,\frac{1}{\gamma^2})}{SN(4K(\frac{1}{\gamma^2})z,\frac{1}{\gamma^2})}, 
\end{equation}
where $SN(\nu,\kappa)$, $CN(\nu,\kappa)$, $DN(\nu,\kappa)$, and $K(\kappa)$ are the Jacobi elliptic sine function, Jacobi elliptic cosine function, Jacobi elliptic delta function and complete elliptic integral of the first kind, respectively.

In the $D\bar{D}$ center-of-mass frame, the momenta $k_i$ is related to $q_i$ as
\begin{equation}
    k_i=\frac{\sqrt{(s-\varepsilon_i^2)(s-\Delta m^{i2})}}{2\sqrt{s}}\cong \sqrt{\frac{\mu_i}{\epsilon_i}}q_i
\end{equation}
with $\mu_i\equiv\frac{m_{D^i}m_{\bar{D}^i}}{m_{D^i}+m_{\bar{D}^i}}$ the reduced mass and $\Delta m^i\equiv m_{D^i}-m_{\bar{D}^i}$ the mass difference of the $i$th channel. In this case, the 
scattering amplitude can be represented as a single-valued function of $z$ with two periods $1$ and $i\tau\equiv i\frac{K(1-\frac{1}{\gamma^2})}{2K(\frac{1}{\gamma^2})}$ 
 by replacing $k_i$ by $q_i$.
 In this case, the eight RSs 
\[
\begin{aligned}
\text{RS}_{+++}: &\quad \operatorname{Im}[q_1] > 0, \quad \operatorname{Im}[q_2] > 0, \quad \operatorname{Im}[q_3] > 0, \\
\text{RS}_{++-}: &\quad \operatorname{Im}[q_1] > 0, \quad \operatorname{Im}[q_2] > 0, \quad \operatorname{Im}[q_3] < 0, \\
\text{RS}_{+-+}: &\quad \operatorname{Im}[q_1] > 0, \quad \operatorname{Im}[q_2] < 0, \quad \operatorname{Im}[q_3] > 0, \\
\text{RS}_{+--}: &\quad \operatorname{Im}[q_1] > 0, \quad \operatorname{Im}[q_2] < 0, \quad \operatorname{Im}[q_3] < 0, \\
\text{RS}_{-++}: &\quad \operatorname{Im}[q_1] < 0, \quad \operatorname{Im}[q_2] > 0, \quad \operatorname{Im}[q_3] > 0, \\
\text{RS}_{-+-}: &\quad \operatorname{Im}[q_1] < 0, \quad \operatorname{Im}[q_2] > 0, \quad \operatorname{Im}[q_3] < 0, \\
\text{RS}_{--+}: &\quad \operatorname{Im}[q_1] < 0, \quad \operatorname{Im}[q_2] < 0, \quad \operatorname{Im}[q_3] > 0, \\
\text{RS}_{---}: &\quad \operatorname{Im}[q_1] < 0, \quad \operatorname{Im}[q_2] < 0, \quad \operatorname{Im}[q_3] < 0.
\end{aligned}
\] 
on the complex $\sqrt{s}$ plane can be conformed onto the $z$-plane as shown by Fig.~\ref{fig:pole position}. 

Due to the doubly periodic property of the Jacobi functions, the scattering amplitude has the period $1$ along the real axis and the period $i\tau$ along the imaginary axis. Through the definition of the ``momentum" $q_i\equiv \sqrt{s-\epsilon_{i}^{2}}$, we can also clarify the position of the $i$th with $i=1,2,3$ threshold by solving equation $q_i(z)=0$. The physical axis is constructed numerically by sampling $\sqrt{s}$ at discrete points in the range $[-1000, 1000]~ \mathrm{GeV}$. The details can be found in the Appendix~\ref{app:z fun}. 
In this way, we eventually plot all the eight RSs onto the complex $z$-plane, enabling clear visualization of the positions of thresholds, the physical axis, pole positions and their trajectories as the coupling constants $g_l$ approach zero, as illustrated in Fig.~\ref{fig:pole position}.

For $l=0$, i.e., Fig.~\ref{fig:pole position} (a), the blue, green and black triangles are the three poles $3.525 - 0.007i$, $3.913-0.016i$, $4.106-0.092i$ for $l=0$ in Table.~\ref{tab:pole_min}. When the couplings tend to zeros, the later two poles come back to the bare state. Only the first one is a dynamically generated state. The black triangle for $l=1$ correspond to the pole $3.761-0.006i$, and it is a renormalized states.

\section{Summary}
\label{sec:Summary}
We investigate the crucial role of three-body final state interactions in the $B^+\to D^0\bar{D}^0K^+$, $B^+\to D^+D^-K^+$, and $B^+\to D_s^+D_s^-K^+$ decays to precisely determine the resonance  parameters in the two-body subsystems. The three-body final state interaction is incorporated by Khuri-Treiman formalism. Its dispersive representation is used, projecting singularities from other channels onto the complex plane of the interested channel. The $D\bar{D}$ scattering amplitude is a key input for the KT equation, which is constructed within HQSS. 
The potential includes both a contact term and contribution from bare charmonium states. With this potential, one can obtain the $D\bar{D}$ scattering amplitude by  solving the LSE to obtain the full scattering amplitude. 
The scheme with two bare states and the three-body final state interaction can describe the experimental data of all the three channels. With the parameters fitted to the data, we find several poles on various Riemann sheets. Two poles, i.e. the pole $3.913-0.016i~\mathrm{GeV}$ for $S$-wave interaction and the pole $3.761-0.006i~\mathrm{GeV}$ for the $P$-wave interaction, have significant contributions to the physical observables. They are identified as $\chi_{c0}(3930)$ and $\psi(3770)$, respectively. To distinguish dynamically generated states from those originating from the input bare states, we track the movement of these poles as the bare state coupling constants to zero. The analysis indicates that the above mentioned poles are indeed evolved from the input bare states. This kind of analysis gives a universal description of the $B^+\to D^0\bar{D}^0K^+$, $B^+\to D^+D^-K^+$, and $B^+\to D_s^+D_s^-K^+$ processes. As it considers the three-body interaction, it can extract the resonance parameters more precisely in the above three channels.

\vspace{0.2cm}
{\bf \color{gray}Acknowledgements:}~~
We are grateful to Yun-Hua Chen and 
Meng-Lin Du for the helpful discussion. 
This work is partly supported by the National Natural Science Foundation of China with Grants No.~12375073, No.~12547105 and National
College Students’ Innovation Training Program (Grant No.~202423011). J.H is a co-first author.

\bibliography{ref.bib}

\onecolumngrid
\newpage
\appendix

\clearpage

\section{THE VALUES OF THE FITTED PARAMETERS}
\label{app:fit parameters}
The values of the fitted parameters are listed in Table~\ref{fit1_para}.  They are divided into three categories: one category is related to the $D\bar{D}$ scattering parameters; another category consists of the subtraction constants in the dispersion relation or in the production vertex; and the third category comprises the normalization constants from differential width distributions to event distributions. 

For the first part, both the contact potential and the bare state contribute to the $D\bar{D}$ scattering. For the former one, there are both $S$-wave and $P$-wave contributions which are encoded in the $V^{0}_{ij}$ and $V^{1}_{ij}$ potentials, respectively. Each part contains 6 parameters. 
The model incorporates two bare states, each defined by three parameters: its bare mass, width, and coupling to $D\bar{D}$. The parameters for the $J^{PC}=0^{++}$ state are $m_0$, $\Gamma_0$, and $g_0$, while those for the $J^{PC}=1^{--}$ state are $m_1$, $\Gamma_1$, and $g_1$. The parameters $r_1$ and $r_2$ are introduced to account for $\mathrm{SU}(3)_f$ breaking. Specifically, the coupling to the $D_s^+D_s^-$ channel is modified by a factor of $r_1$ for processes involving a single $D_s^+D_s^-$ pair, and by $r_2$ for processes involving two such pairs. As the result, the $S$-wave potential between $D^i\bar{D}^i$ and $D^j\bar{D}^j$ is written as,
\begin{equation}
    (V_{\text{eff}}^{S})_{ij}=V_{ij}^{0}+\frac{g_0^2}{s-m_{0}^2+im_0\Gamma_{0}}.
\end{equation}
for $i,j=1,2$ channels. 
For $i=3,j=1,2$ or $i=1,2, j=3$, the potential is represented as
\begin{equation}
    (V_{\text{eff}}^{S})_{ij}=V_{ij}^{0}+\frac{r_1g_0^2}{s-m_{0}^2+im_0\Gamma_{0}}.
\end{equation}
For $i=j=3$, the potential is given by
\begin{equation}
    (V_{\text{eff}}^{S})_{ij}=V_{ij}^{0}+\frac{r_2^2g_0^2}{s-m_{0}^2+im_0\Gamma_{0}}.
\end{equation}

Except for the parameters in the potential, there is also one additional parameter $\Lambda$ in the exponent form factor $\mathrm{exp}(-2\mathbf{q}^2/\Lambda^2)$ which is used to make the two-body propagator 
\begin{equation}
    G_{i}^{l}(s)=\int \frac{d^3q}{(2\pi)^3}\frac{\mathbf{q}^{2l}\mathrm{exp}(-2\mathbf{q^2/\Lambda^2})}{\sqrt{s}-m_{D^i}-m_{\overline{D}^i}-\mathbf{q}^2/2\mu_i}
\end{equation}
convergent. 

The treatment of three-body final state interactions are different for the two fitting schemes. Scheme I incorporates three-body final state interactions, including the contribution of the $X_0(2900)$ resonance. The coupling $g$ is a free parameter, describing the coupling between the $X_0(2900)$ and the $D^-K^+$ channel. The influence of the $K^+$ on the $D\bar{D}$ rescattering is included via dispersion relations. A first subtraction is applied to ensure the convergence as $s\to \infty$ which necessitates six subtraction constants, i.e. three subtraction constants $C_{0,i}$ with $i=1,2,3$ for the $S$-wave channels and three subtraction constants $C_{1,i}$ with $i=1,2,3$ for the $P$-wave channels. 
Scheme II offers an alternative interpretation, where the subtraction constants represent the direct production vertices of the $B^+\to D\bar{D}K^+$ processes. This scheme does not include the $X_0(2900)$ contribution, consequently the parameter $g$ is not present. 

As the experimental data are the event distributions instead of the partial width distributions, we introduce three normalization constants $a_i$ with $i=1,2,3$ 
\begin{equation}
    \frac{dN_i}{d\sqrt{s}}=a_i\frac{d\Gamma_i}{d\sqrt{s}},
\end{equation}
for the three channels.
\newpage
\begin{table}[h!]
        \caption{The fitted parameters of scheme I and scheme II are listed in this table. N/A represents that the parameter $g$ does not appear in scheme II. The numbers in the bracket are the degrees of freedom. The parameters in the first block are for the $D_{(s)}\bar{D}_{(s)}$ scattering. The last three parameters define the normalization that converts the partial width into the event distribution.}
    \centering
    \begin{tabular}{c@{\hspace{60pt}}c@{\hspace{60pt}}c}
       \hline
        Parameters& Scheme I & Scheme II\\
        \hline
        $V^{0}_{11}\ [\mathrm{GeV}^{-2}]$ &  $-1164.165\pm 0.290$  & $-1209.392\pm1.813$ \\
        $V^{0}_{12}\ [\mathrm{GeV}^{-2}]$ &  $-1213.237\pm 0.245$   &  $-1215.689\pm 1.352$\\
        $V^{0}_{13}\ [\mathrm{GeV}^{-2}]$ &  $-52.593\pm 0.871$  & $-84.585\pm0.476$ \\
        $V^{0}_{22}\ [\mathrm{GeV}^{-2}]$ &  $-1264.547\pm 0.192$  &  $-1230.654\pm2.443$\\
        $V^{0}_{23}\ [\mathrm{GeV}^{-2}]$ &  $-53.940\pm 0.882$  & $-84.926\pm0.479$ \\
        $V^{0}_{33}\ [\mathrm{GeV}^{-2}]$ &  $-24.885\pm 1.274$  &  $-43.739\pm2.518$\\
        $V^{1}_{11}\ [\mathrm{GeV}^{-2}]$ &  $-195.854\pm 22.321$  &  $-174.847\pm 11.990$\\
        $V^{1}_{12}\ [\mathrm{GeV}^{-2}]$ &  $1408.859\pm 156.241$   &  $-806.245\pm 38.692$\\
        $V^{1}_{13}\ [\mathrm{GeV}^{-2}]$ &  $-326.275\pm 24.731$  &  $267.703\pm 20.065$\\
        $V^{1}_{22}\ [\mathrm{GeV}^{-2}]$ &  $-14657.093\pm 69.067$  &  $-5860.571\pm 126.068$\\
        $V^{1}_{23}\ [\mathrm{GeV}^{-2}]$ &  $3005.169\pm 17.778$  &  $2996.321\pm 33.930$\\
        $V^{1}_{33}\ [\mathrm{GeV}^{-2}]$ &  $-637.008\pm 9.413$  &  $-1566.025\pm 34.887$\\
        $m_0\ [\mathrm{GeV}]$ & $3.909\pm 0.002$   &  $3.892\pm 0.003$\\
        $\Gamma_0\  [\mathrm{GeV}]$ & $0.035\pm 0.006$   &  $0.029\pm0.005$\\
        $g_0\ [\mathrm{GeV}^{0}]$ &  $12.409\pm 0.537$  &  $22.126\pm0.600$\\
        $m_1\ [\mathrm{GeV}]$ &  $3.802\pm 0.006$  &  $3.803\pm 0.005$\\
        $\Gamma_1\  [\mathrm{GeV}]$ & $0.013\pm 0.004$   &  $0.018\pm 0.005$\\
        $g_1\ [\mathrm{GeV}^{0}]$ &  $17.465\pm1.663$  &  $30.000\pm23.011$\\
        $r_1\ [\mathrm{GeV}^{0}]$ &$0.462\pm0.009$ & $0.353\pm 0.012$\\
        $r_2\ [\mathrm{GeV}^{0}]$&$0.235\pm 0.011$ & $0.447\pm 1.249$\\
        $\Lambda\ [\mathrm{GeV}]$&$1.591\pm0.041$ & $1.226 \pm 0.024$\\\hline
        $g\ [\mathrm{GeV}^0]$&$26.994\pm1.264$ & N/A\\
        $c_{0,1}\ [\mathrm{GeV}^{0}]$ & $-84496.268\pm137.510$& $-6284.299\pm 805.688$\\
        $c_{0,2}\ [\mathrm{GeV}^{0}]$ & $79960.286\pm 137.431$& $4213.646\pm 789.301$\\
        $c_{0,3}\ [\mathrm{GeV}^{0}]$ & $-4490.900\pm642.328$& $-56470.539\pm 7301.609$\\
        $c_{1,1}\ [\mathrm{GeV}^{0}]$ & $95.387\pm 55.748$& $313.538\pm 54.324$\\
        $c_{1,2}\ [\mathrm{GeV}^{0}]$ &$148.855\pm 13.554$ & $91.776\pm51.268$\\
        $c_{1,3}\ [\mathrm{GeV}^{0}]$ & $1228.214\pm101.298$& $2377.140\pm330.626$\\\hline
        $a_1\ [\mathrm{GeV}^{-1}]$& $0.153\pm 0.020$& $0.064\pm 0.016$\\
        $a_2\ [\mathrm{GeV}^{-1}]$& $3.281\pm 0.433$& $0.248\pm 0.056$\\
        $a_3\ [\mathrm{GeV}^{-1}]$& $0.010\pm 0.001$& $0.0003\pm 0.00007$\\\hline
        $\chi^2/(\mathrm{d.o.f})$ & 2.488(389)& 1.758(100)\\
        \hline
    \end{tabular}
    \label{fit1_para}
\end{table}

\section{THE $\Lambda$ DEPENDENCE OF THE RESULTS}
\label{app:B}
To check the stability of our results, we vary the values of $\Lambda$ as $1.432~\mathrm{GeV},~1.591~\mathrm{GeV},~1.750~\mathrm{GeV},~1.909~\mathrm{GeV},~2.068~\mathrm{GeV}$ and refit the experimental data. For the above five values, the reduced chi squares are $2.568$, $2.488$, $2.543$, $2.627$ and $2.719$, respectively. 
The corresponding invariant mass distributions with those five $\Lambda$s are plotted in Fig.~\ref{fig:diff lambda} as orange, cyan, green, brown, and purple solid lines in order. One can see that both the reduced chi squares and the invariant mass distributions are almost the same. We also check the pole positions listed in Table.~\ref{tab:pole_0} and Table~\ref{tab:pole_1} for various $\Lambda$s. The two tables show that the pole positions on various RSs are also stable, which is a natural result, as the pole is driven by the experimental data and our framework can describe the experimental data well. In this case, we conclude that our results are insensitive to the parameter $\Lambda$.

\begin{table}[h!]
    \centering
    \renewcommand{\arraystretch}{1.3}
    \caption{The poles for $l=0$ on various RSs.}
    \resizebox{0.9\textwidth}{!}{
    \begin{tabular}{|c|c|c|c|c|c|}
    \hline
        RS &$\Lambda=1.432$& $\Lambda=1.591$& $\Lambda=1.750$ & $\Lambda=1.909$& $\Lambda=2.068$\\
        \hline
       $+++$ & $3.524-0.007i$ & $3.527-0.007i$& $3.519-0.007i$&$3.491-0.007i$ & $3.455-0.006i$\\
         \hline
      $--+$ & $3.913-0.016i$ & $3.911-0.017i$ &$3.910-0.018i$ & $3.908-0.018i$ & $3.906-0.017i$\\
      \hline
      $---$ & $4.112-0.087i$ & $4.106-0.092i$ & $4.100-0.093i$ &$4.092-0.089i$  & $4.090-0.086i$\\
      \hline
    \end{tabular}
    }
    \renewcommand{\arraystretch}{\origArraystretch}
    \label{tab:pole_0}
\end{table}

\begin{table}[h!]
    \centering
    \renewcommand{\arraystretch}{1.3}
    \caption{The poles for $l=1$ on various RSs.}
    \resizebox{0.9\textwidth}{!}{
    \begin{tabular}{|c|c|c|c|c|c|}
    \hline
        RS & $\Lambda=1.432$& $\Lambda=1.591$& $\Lambda=1.750$ & $\Lambda=1.909$& $\Lambda=2.068$\\
        \hline
      $--+$ & $3.772-0.006i$ &$3.771-0.006i$ & $3.771-0.005i$& $3.767-0.004i$& $3.765-0.002i$\\
      \hline
    \end{tabular}
    }
    \renewcommand{\arraystretch}{\origArraystretch}
    \label{tab:pole_1}
\end{table}

\begin{figure*}[ht!]
\centering
    \includegraphics[width=1\textwidth]{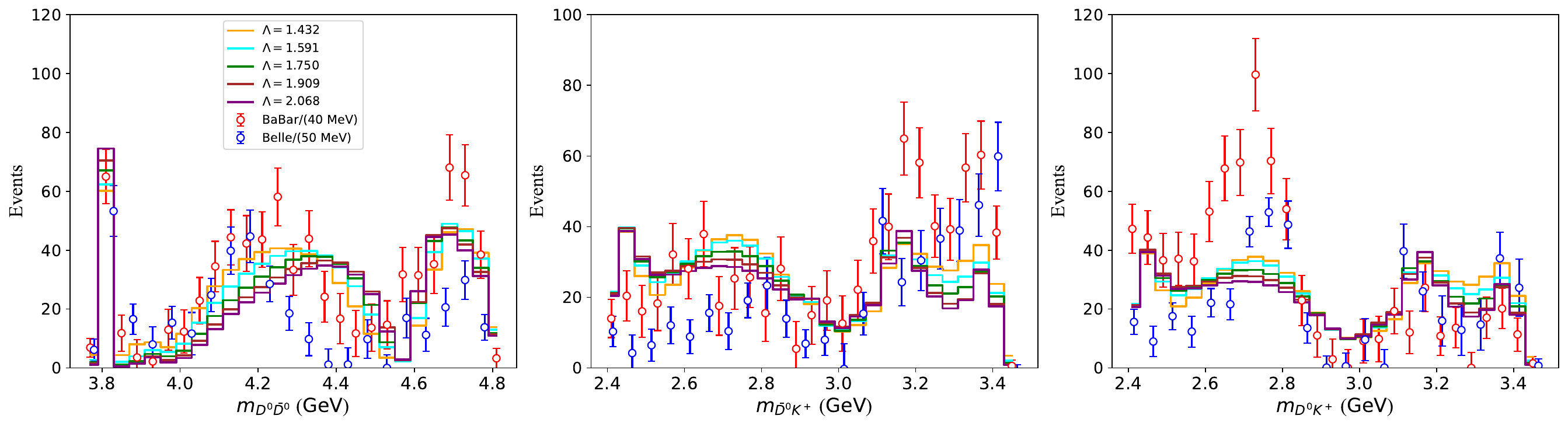}
    \includegraphics[width=1\textwidth]{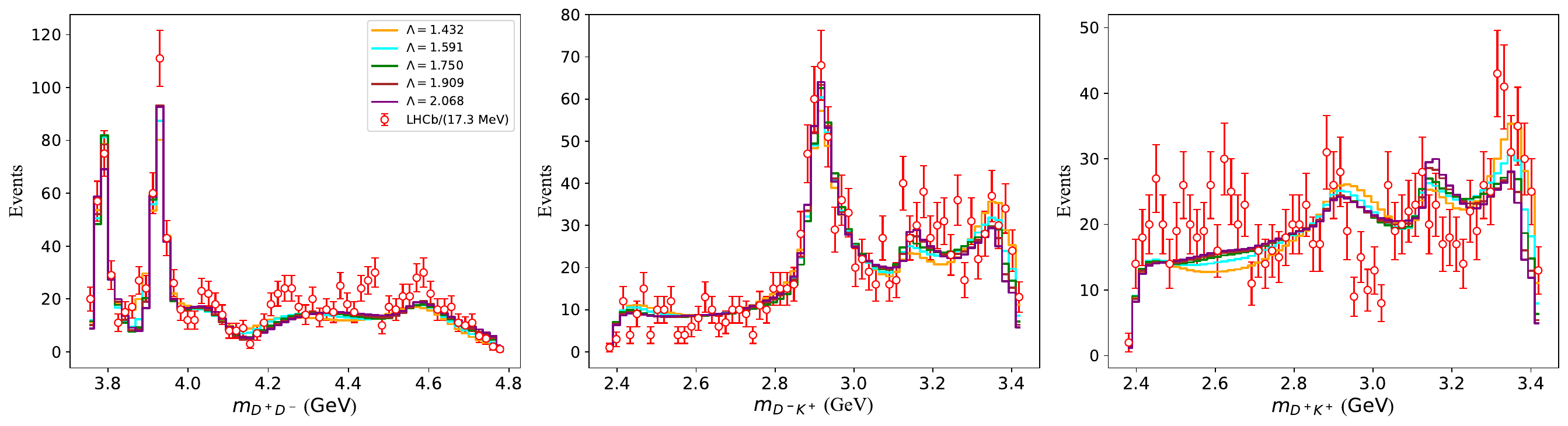}
    \includegraphics[width=1\textwidth]{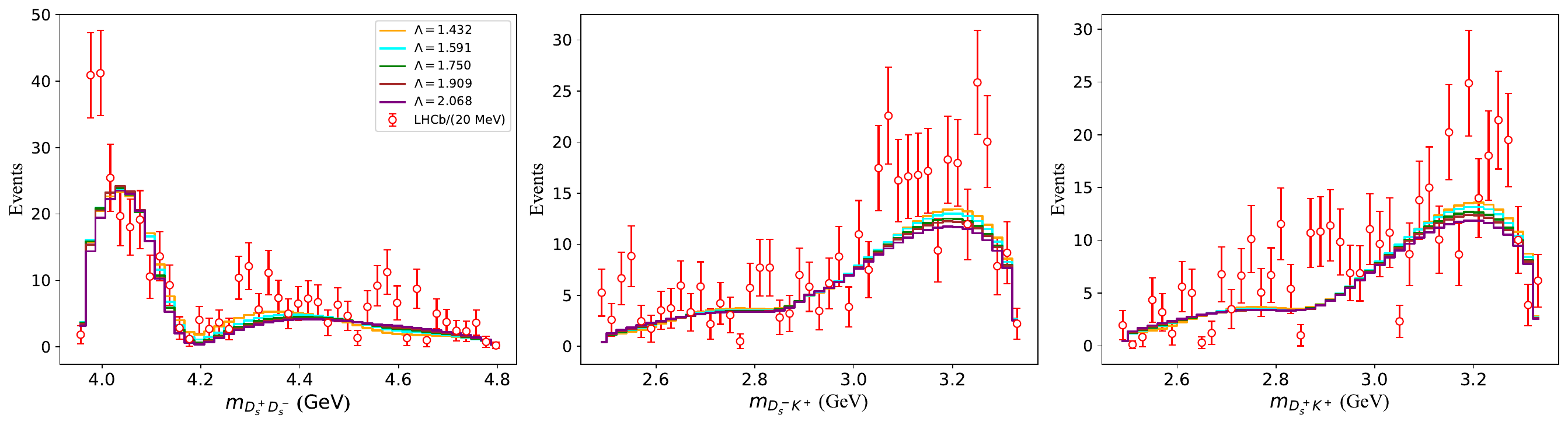}
    \caption{The invariant mass spectra of $D\overline{D}$, $\overline{D}K$ and $DK$. The first line are the  fit results of $B^+\to D^0\overline{D}^0K^+$ process, where we plot the experiment results of both $BABAR$ (red) and Belle (blue); The second line are the results of $B^+\to D^+D^-K^+$ process with the experimental data (red hollow points) from LHCb collaboration; The third line are the results of $B^+\to D_s^+D_s^-K^+$ process, with the experimental data (red hollow points) from LHCb collaboration. The orange, cyan, green, brown, purple solid step lines represent the fit results for $\Lambda$ values of $1.432$, $1.591$, $1.750$, $1.909$, and $2.068$ , respectively.}
    \label{fig:diff lambda}
\end{figure*}

\section{The COMPOSITENESS OF THREE-CHANNEL CASE FOR $S$-WAVE INTERACTION}
\label{app: C}
The standard Effective-Range-Expansion (ERE) of the $S$-wave scattering amplitude reads as 
\begin{eqnarray}
    t^{0}(s)=-\frac{2\pi}{\mu}\frac{1}{-1/a_0+1/2r_0k^2-ik+\mathcal{O}(k^4)},
\end{eqnarray}
where $s=(m_1+m_2+\frac{k^2}{2\mu})^2$, and $\mu=\frac{m_1m_2}{m_1+m_2}$ is the reduced mass of scattering system. $a_0$ and $r_0$ are scattering length and effective range, respectively, which can be obtained from the following equations
\begin{equation}
    \begin{aligned}
        &a_0=\left. \frac{\mu}{2\pi}t^{0}(s)\right|_{s\to s_{th}},\\
        &r_0=-\left. \frac{2\pi}{\mu}\mathrm{Re}\left[\frac{d(t^{0}(\sqrt{s}))^{-1}}{d\sqrt{s}}\right]\right|_{\sqrt{s}\to\sqrt{s_{th}}}.
    \end{aligned}
\end{equation}
However, as discussed in Ref.~\cite{Baru:2021ldu}, the effective range can gain the corrections from the coupled channels. To obtain the compositeness of a given particle, one should remove those corrections. As our case is very general for the three-coupled channel case, we would propose a general solution to remove the correction from both isospin and SU(3) breaking effects. The coupled-channel corrections come from the threshold difference of the couple channels. 
We can extract these corrections from the difference between the full scattering amplitude and that of the SU(3) limit. 
To illustrate this, we take two-channel case as an example. 
The potential of the two-channel case is a two-dimensional symmetry matrix
\begin{equation}
    V(E)=\begin{pmatrix}
        v_{11}&v_{12}\\
        v_{12}&v_{22}
    \end{pmatrix},
\end{equation}
which could be energy dependent 
with $E=\sqrt{s}-m_1-m_2$ is the center-of-mass energy relative to the first threshold. The two-point correlation function is a diagonal matrix
\begin{equation}
    G(E)=\begin{pmatrix}
        G(E)&0\\
        0&G(E-\Delta)
    \end{pmatrix}.
\end{equation}
with $\Delta$ is the energy gap between the second and the first threshold. With the above formulas, one can extract the scattering amplitude from the first channel to itself as 
\begin{equation}
    t_{11}^{-1}(E)=\frac{1}{v_{11}}-G(E)-\frac{v_{12}^2G(E-\Delta)}{v_{11}[v_{11}-G(E-\Delta)(v_{12}^2-v_{11}v_{22})]}.
    \label{eq: app_t11}
\end{equation}
From Eq.~\eqref{eq: app_t11}, the effect of the coupled channel effect comes from the third term. We can extract the corrections from the coupled channel case by comparing Eq.~\eqref{eq: app_t11} and $t_{11}^{-1}(E)$ at the $\Delta=0$ limit, which read as 
\begin{equation}
    \Delta r=-\frac{2\pi}{\mu}\mathrm{Re}\left.\left[ \frac{d(t^{-1}_{11}-t^{-1}_{11}|_{\Delta\to0})}{dE}\right]\right|_{E\to 0}.
\end{equation}
The effective range after removing the coupled-channel correction read as 
\begin{equation}
    r^{\prime}=r-\Delta r.
\end{equation}
Furthermore, the Weinberg criterion, which is typically used to characterize the compositeness of bound, virtual and resonance states, which can be found in Refs.~\cite{Baru:2021ldu,Niu:2024cfn,Matuschek:2020gqe,Li:2021cue},
\begin{eqnarray}
    \bar{X}=\frac{1}{\sqrt{1+2|\frac{r^{\prime}}{\mathrm{Re}[a_0]}|}}.
\end{eqnarray}

\section{ANALYTIC MAP OF THREE-CHANNEL $S$-MATRIX}
\label{app:z fun}
As a function of $\sqrt{s}$, the Riemann surface for a three-channel $S$-matrix is an eight-sheeted complex plane with three branch points at $\sqrt{s}=\epsilon_1,\epsilon_2$and $\epsilon_3$($\epsilon_1<\epsilon_2<\epsilon_3$), where $\epsilon_i=M_i+M_{i}^{'}$ is the threshold energy, $M_i$ and $M_{i}^{'}$ are the masses of DD meson in the $i$th channel. Following \cite{Yamada:2022xam}, we can map the eight Riemann sheets of the complex $\sqrt{s}$ plane onto a torus by constructing a three-channel uniformization variable $z$.

We first define the momentum, $q_i=\sqrt{s-\epsilon_{i}^{2}}$, and $\gamma=\frac{\sqrt{\epsilon_{3}^{2}-\epsilon_{1}^{2}}+\sqrt{\epsilon_{3}^{2}-\epsilon_{2}^{2}}}{\sqrt{\epsilon_{2}^{2}-\epsilon_{1}^{2}}}$ for later use. The first step is the same as the uniformization of the two-channel S matrix for channels 1 and 2. We define $z_{12}$ by
\begin{equation}
    z_{12}=\frac{q_1+q_2}{\sqrt{\epsilon_2^2-\epsilon_1^2}}
\end{equation}
Then $q_1$, $q_2$ and $q_3$ are represented by $z_{12}$
\begin{equation}
    \begin{aligned}
    q_1&=\frac{\sqrt{\epsilon_2^2-\epsilon_1^2}}{2}(z_{12}+\frac{1}{z_{12}}),\\
    q_2&=\frac{\sqrt{\epsilon_2^2-\epsilon_1^2}}{2}(z_{12}-\frac{1}{z_{12}}),\\
    q_3&=\frac{\sqrt{\epsilon_2^2-\epsilon_1^2}}{2}z_{12}\sqrt{(1-\frac{\gamma^2}{z_{12}^2})(1-\frac{1}{\gamma^2z_{12}^2})}
    \label{q(z12)}
    \end{aligned}
\end{equation}\\
To uniformize the third channel \cite{Yamada:2022xam}, we can define the three-channel uniformization variable $z$ as
\begin{equation}
    z=\frac{1}{4K(\frac{1}{\gamma^2})}SN^{-1}(\frac{\gamma}{z_{12}},\frac{1}{\gamma^2})
    \label{z(z12)}
\end{equation}
where $K(\kappa)$ and $SN^{-1}(\nu,\kappa)$ are the complete elliptic integral of the first kind and the inverse Jacobi’s elliptic function respectively. Inversely, $z_{12}$ is given as a single-valued function of $z$ by
\begin{equation}
    z_{12}=\frac{\gamma}{SN(4K(\frac{1}{\gamma^2})z,\frac{1}{\gamma^2})}
\end{equation}
where $SN(\nu,\kappa)$ is the Jacobi elliptic sine function. Then $q_1$, $q_2$, and $q_3$ are given as single-valued functions of $z$ by
\begin{equation}
    \begin{aligned}
    q_1&=\frac{\sqrt{\varepsilon_2^2-\varepsilon_1^2}}{2}[\frac{\gamma}{SN(4K(\frac{1}{\gamma^2})z,\frac{1}{\gamma^2})}+\frac{SN(4K(\frac{1}{\gamma^2})z,\frac{1}{\gamma^2})}{\gamma}],\\
    q_2&=\frac{\sqrt{\varepsilon_2^2-\varepsilon_1^2}}{2}[\frac{\gamma}{SN(4K(\frac{1}{\gamma^2})z,\frac{1}{\gamma^2})}-\frac{SN(4K(\frac{1}{\gamma^2})z,\frac{1}{\gamma^2})}{\gamma}],\\
    q_3&=\frac{\sqrt{\varepsilon_2^2-\varepsilon_1^2}}{2}\frac{\gamma CN(4K(\frac{1}{\gamma^2})z,\frac{1}{\gamma^2})DN(4K(\frac{1}{\gamma^2})z,\frac{1}{\gamma^2})}{SN(4K(\frac{1}{\gamma^2})z,\frac{1}{\gamma^2})}
    \end{aligned}
\end{equation}
where $CN(\nu,\kappa)$ and $DN(\nu,\kappa)$ are Jacobi elliptic cosine function and Jacobi elliptic delta function,respectively.

The scattering amplitude from the $i$th channel to the $j$th channel is related to the $S$-matrix, as $A_{ij}=\frac{S_{ij}-\delta_{ij}}{2i\sqrt{k_ik_j}}$. In the mass-center system,
\begin{equation}
    k_i=k_i^{'}=\frac{\sqrt{(s-\varepsilon_i^2)(s-\Delta M_i^2)}}{2\sqrt{s}}\cong \sqrt{\frac{\mu_i}{\epsilon_i}}q_i
\end{equation}
where $\mu_i=\frac{M_iM_{i}^{'}}{M_i+M_{i}^{'}}$ is the reduced mass in the $i$th channel. The difference between $k_i$ and $\sqrt{\frac{\mu_i}{\epsilon_i}}q_i$ can be ignored when $\Delta M_i=M_i-M_i^{'}$ is small enough.

In this way, the three-channel scattering amplitude $A$ is given as a single-valued function of $z$ with two periods 1 and $i\tau$, where $\tau=\frac{K(1-\frac{1}{\gamma^2})}{2K(\frac{1}{\gamma^2})}$. So, all the poles of $A(z)$ would be indicated on the complex $z$ plane, which ranged from $-\frac{1}{2}-\frac{1}{2}\tau i$ to $\frac{1}{2}+\frac{1}{2}\tau i$. 
  
To clarify the position of the threshold energy, we can set $q_i=0$, with $i=1,2,3$ because of the definition of momentum, $q_i=\sqrt{s-\epsilon_i^2}$. From Eq.~\eqref{q(z12)}, we can find that $z_{12}=\pm i$, $z_{12}=\pm 1$ and $z_{12}=\pm \gamma, \pm \frac{1}{\gamma}$ correspond to the positions of three threshold energy respectively. Through Eq.~\eqref{z(z12)}, $z_{12}$ can be translated into $z$. In this way, we can plot the positions of the threshold energy on the z-plane.

In the $z_{12}$-plane, the physical region of a real energy E is indicated by a line in Fig.\ref{physical axis}. We can also translate the physical axis in $z_{12}$-plane into the $z$-plane by Eq.~\eqref{z(z12)}, from which we can indicate the physical region of a real energy E in $z$-plane, as shown in Fig.\ref{physical axis}.
  
  \begin{figure}[htp]
    \centering
    \includegraphics[width=1\textwidth]{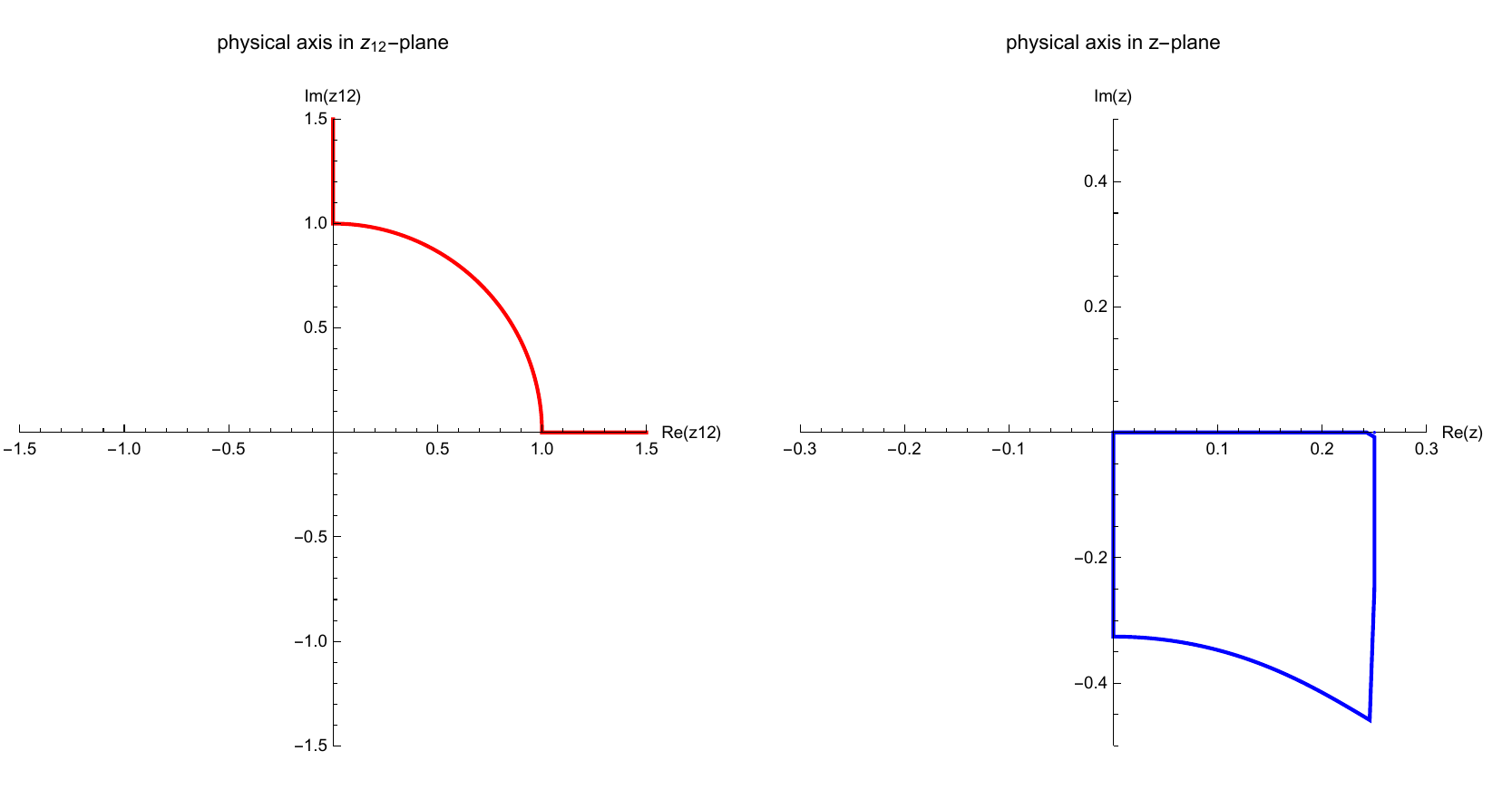}
    \caption{the left one shows the physical axis in $z_{12}$-plane, and the right one shows the physical axis in z-plane}
    \label{physical axis}
\end{figure}
Since a meromorphic function with double periods can be topologically mapped on a torus, we map the function $A(z)$ on the torus so that the position of the poles of $A(z)$ can vary on the torus consistently. The parametric equations of a torus can be written as
  \begin{equation}
      \begin{aligned}
      x&=(R+r\cos{\theta})\cos{\phi}\\
      y&=(R+r\cos{\theta})\sin{\phi}\\
      z&=r\sin{\theta}
      \end{aligned}
  \end{equation}
 which have only two free parameters $\theta$ and $\phi$. Through the transform
  \begin{equation}
      \mathrm{Re}[z]=\frac{1}{4\pi}\phi,     
      \mathrm{Im}[z]=\frac{\tau}{4\pi}\theta
  \end{equation}
we can map the complex $z$ plane ranged from $-\frac{1}{2}-\frac{1}{2}\tau i$ to $\frac{1}{2}+\frac{1}{2}\tau i$ onto a torus. 
The all poles position in our calculate as shown in Table.~\ref{tab:pole_all}.
\begin{table}[h]
    \centering
    \renewcommand{\arraystretch}{1.3}
    \caption{The poles on eight RSs. The real part of $\sqrt{s}$ is the energy and the imaginary part is the half width. }
    \resizebox{0.45\textwidth}{!}{
    \begin{tabular}{|c|c|c|c|}
    \hline
        RS & $l$&$z$&$\sqrt{s}$ \\
        \hline
       $+++$  & $S$-wave& $-0.002-0.138i$& $3.515-0.007i$ \\
      $+++$  & $S$-wave&$-0.229-0.063i$& $3.911-0.017i$\\
      $--+$ & $S$-wave& $0.498-0.136i$& $3.507-0.008i$\\
      $--+$& $S$-wave& $0.271-0.063i$& $3.911-0.017i$\\
      $---$& $S$-wave& $0.229+0.063i$& $3.911-0.017i$\\
      $---$& $S$-wave& $0.125+0.020i$& $4.105-0.091i$\\
      $---$& $S$-wave& $-0.127+0.015i$& $4.105-0.065i$\\
      $++-$& $S$-wave& $0.372+0.015$& $4.105+0.065i$\\
      $++-$& $S$-wave& $-0.375+0.020i$& $4.104-0.091i$\\
      \hline
      $+++$  &$P$-wave &$-0.236-0.241i$ & $3.772-0.006i$ \\
      $-++$  &$P$-wave &$0.000-0.554i$ & $3.591-0.000i$ \\
      $--+$  &$P$-wave &$-0.500-0.166i$ & $3.599-0.000i$ \\
      $--+$  &$P$-wave &$0.266-0.241i$ & $3.772-0.006i$ \\
      $---$  &$P$-wave &$0.001+0.207$ & $3.663-0.001i$ \\
      $---$  &$P$-wave &$0.000+0.166i$ & $3.599-0.000i$ \\
      \hline
    \end{tabular}
    }
    \renewcommand{\arraystretch}{\origArraystretch}
    \label{tab:pole_all}
\end{table}
Fig.~\ref{trajectories on torus} shows the trajectories on torus of the $S$-wave and the $P$-wave poles as the coupling constants $g_l$ approach zero. 
\begin{figure*}[htp]
    \centering
     \subfigure[The trajectories of S wave.]{
    \includegraphics[width=0.48\textwidth]{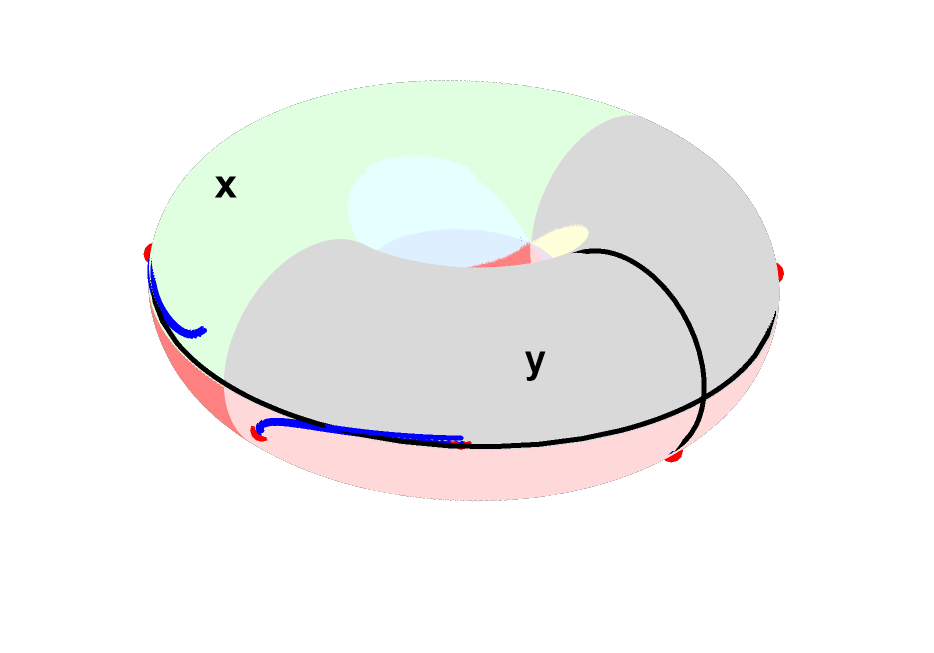}
    \label{trajectories of s wave on torus}
    }
    \subfigure[The trajectories of P wave.]{
    \includegraphics[width=0.48\textwidth]{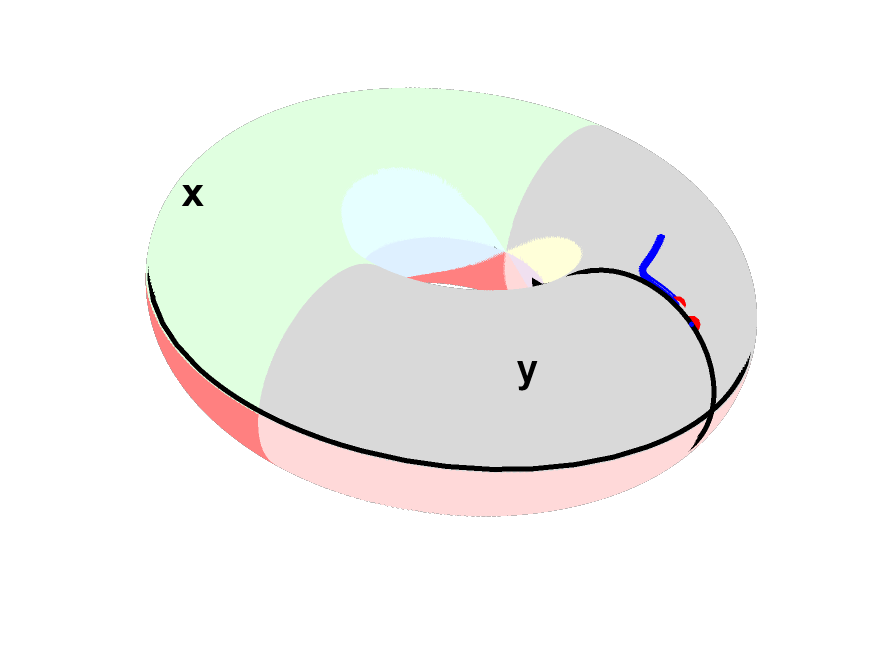}
    \label{trajectories of p wave on torus}
    }
    \includegraphics[width=0.32\textwidth]{figure/legendre.pdf}
    \caption{
The initial pole positions are marked by red dots, with their trajectories indicated by blue curves. 
The black lines represent the real (the $x$-axis) and imaginary (the $y$-axis) axes, where the imaginary axis penetrates the torus. 
Colored regions on the torus correspond to distinct Riemann sheets, with the color-sheet mapping given in the legend.
}
    \label{trajectories on torus}
\end{figure*}

\end{document}